\pdfoutput=1
\RequirePackage{ifpdf}
\ifpdf 
\documentclass[pdftex]{sigma}
\else
\documentclass{sigma}
\fi

\usepackage{cite}
\numberwithin{equation}{section}
\numberwithin{remark}{section}

\begin{document}

\allowdisplaybreaks

\renewcommand{\thefootnote}{$\star$}

\renewcommand{\PaperNumber}{066}

\FirstPageHeading

\ShortArticleName{A New Class of Solvable Many-Body Problems}

\ArticleName{A New Class of Solvable Many-Body Problems\footnote{This
paper is a contribution to the Special Issue ``Superintegrability, Exact Solvability, and Special Functions''. The full collection is available at \href{http://www.emis.de/journals/SIGMA/SESSF2012.html}{http://www.emis.de/journals/SIGMA/SESSF2012.html}}}

\Author{Francesco CALOGERO and Ge YI}

\AuthorNameForHeading{F.~Calogero and G.~Yi}

\Address{Physics Department, University of Rome ``La Sapienza'', \\
Istituto Nazionale di Fisica Nucleare, Sezione di Roma, Italy}
\Email{\href{mailto:francesco.calogero@roma1.infn.it}{francesco.calogero@roma1.infn.it}, \href{mailto:francesco.calogero@uniroma1.it}{francesco.calogero@uniroma1.it},\\
 \hspace*{13.5mm}\href{mailto:ge.yi@roma1.infn.it}{ge.yi@roma1.infn.it}, \href{mailto:yigeking@mail.ustc.edu.cn}{yigeking@mail.ustc.edu.cn}}

\ArticleDates{Received June 27, 2012, in f\/inal form September 20, 2012; Published online October 02, 2012}

\Abstract{A \textit{new} class of \textit{solvable} $N$-body problems is identif\/ied.
They describe $N$ unit-mass point particles whose time-evolution, generally
taking place in the \textit{complex} plane, is characterized by \textit{Newtonian} equations of motion ``of goldf\/ish type'' (acceleration equal
force, with specif\/ic velocity-dependent one-body and two-body forces)
featuring several arbitrary coupling constants. The corresponding
initial-value problems are solved by f\/inding the eigenvalues of a
time-dependent $N\times N$ matrix $U(t)$ explicitly def\/ined in
terms of the initial positions and velocities of the $N$ particles. Some of
these models are \textit{asymptotically isochronous}, i.e.\ in the remote
future they become completely periodic with a period $T$ independent of the
initial data (up to exponentially vanishing corrections). Alternative
formulations of these models, obtained by changing the dependent variables
from the $N$ zeros of a monic polynomial of degree $N$ to its $N$
coef\/f\/icients, are also exhibited.}


\Classification{70F10; 70H06; 37J35; 37K10}

\renewcommand{\thefootnote}{\arabic{footnote}}
\setcounter{footnote}{0}

\section{Introduction}

In this paper a \textit{new} class of \textit{solvable} $N$-body problems is
identif\/ied. They describe an arbitrary number $N$ of unit-mass point
particles whose time-evolution, generally taking place in the \textit{comp\-lex} plane, is characterized by \textit{Newtonian} equations of motion ``of
goldf\/ish type'' (acceleration equal force, with specif\/ic velocity-dependent
one-body and two-body forces, see below) featuring several arbitrary
coupling constants. The \textit{solvable} character of these models is
demonstrated by the possibility to ascertain their time evolution by purely
\textit{algebraic} operations. In particular it is shown below that their
initial-value problems are solved by f\/inding the eigenvalues of a
time-dependent $N\times N$ matrix $U(t) $ explicitly def\/ined in
terms of the initial positions and velocities of the $N$ particles. Some of
these models are \textit{asymptotically isochronous}, i.e.\ in the remote
future they become completely periodic with a period $T$ independent of the
initial data (up to exponentially vanishing corrections). Alternative
formulations of these models, obtained by changing the dependent variables
from the $N$ zeros of a monic polynomial of degree $N$ to its~$N$
coef\/f\/icients, are also exhibited.

The main idea to obtain these results is to identify the $N$ coordinates $%
z_{n}(t) $ characterizing the positions of the particles of the $%
N$-body problem with the $N$ eigenvalues of an $N\times N$ matrix $U(t) $, itself evolving according to a \textit{solvable} (or \textit{integrable}) matrix ODE. This technique was invented long ago by
Olshanetsky and Perelomov \cite{OP1, OP2, OP3, OP4} and has been much
exploited subsequently, identifying thereby many \textit{solvable} (or
\textit{\ integrable}) many-body problems: see for instance \cite{C2001} (in
particular its Section~2.1.3.2 entitled ``The technique of solution of
Olshanetsky and Perelomov (OP)'') and \cite{C2008} (in particular its Section
4.2.2 entitled ``Goldf\/ishing''), and the more recent papers \cite{CC1, CC2,
CC3, CC4, CC5, CC6, CC7, CC8, CC9, CC10, CC11, CC12, CC13, CC14, CC15, CC16,
CC17}. The present paper provides \textit{new} results of this kind, by
taking as point of departure a \textit{solvable} ODE characterizing the
time-evolution of the $N\times N$ matrix $U(t) $ which is
dif\/ferent or more general than those previously employed to this end. These
results are treated in the following section, including its subsections and
 Appendices~\ref{appendixA} and~\ref{appendixB}  where all the equations solved in this paper are
listed (hence, the reader wishing to get an immediate glance at them may
immediately jump to these appendices). The last section, entitled ``Outlook'',
tersely outlines further developments whose treatment is postponed to
subsequent papers.

\section[\textit{Solvable} $N$-body problems]{\textit{Solvable} $\boldsymbol{N}$-body problems}\label{section2}

In this section we describe two classes of \textit{solvable} $N$-body
problems. The models of the f\/irst class are not new, hence their treatment
is not elaborated beyond their identif\/ication; several models of the second
class are \textit{new,} hence they are fully dealt with.

In Subsection~\ref{section2.1} we introduce a system of two matrix f\/irst-order ordinary dif\/ferential equations (ODEs)
characterizing the time-evolution of the
two $N\times N$ matrices $U\equiv U(t) $ and $V\equiv V (
t ) $, and we indicate how the corresponding initial-value problem can
be explicitly solved.

In Subsection~\ref{section2.2} we show how~-- via the introduction of two appropriate
\textit{ans\"{a}tze}~-- the $N$ eigenvalues $z_{n}(t) $ of the
matrix $U(t) $ can be identif\/ied with the $N$ coordinates of $N$
unit-mass point particles whose time-evolution, generally taking place in
the \textit{complex} plane, is characterized by \textit{Newtonian} equations
of motion (``acceleration equal force'', with nonlinear one-body and two-body
forces). These models are thereby shown to be \textit{solvable}. The f\/irst
\textit{ansatz} yields various models whose solvability was already known;
hence their treatment is limited to the derivation of their equations of
motion. The second \textit{ansatz} yields several \textit{new} models~-- as
well as several previously known models~-- characterized by equations of
motion with specif\/ic velocity-dependent one-body and two-body forces ``of
goldf\/ish type'' featuring several arbitrary coupling constants. They are all
listed in  Appendix~\ref{appendixA}. The alternative class of many-body models
obtained by changing the dependent variables from the $N$ zeros of a monic
polynomial of degree $N$ to its $N$ coef\/f\/icients is discussed in Subsection~\ref{section2.3}; the corresponding equations of motion are listed
in  Appendix~\ref{appendixB}.

As it is well known (see for instance \cite{C2001}, in particular Chapter~4,
entitled ``Solvable and/or integrable many-body problems in the plane,
obtained by complexif\/ication''), models such as those described below, which
describe the motions of $N$ points in the \textit{complex} $z$-plane, can be
reformulated as $N$-body models describing the motion in a plane of $N$
point-particles, the positions of which are identif\/ied by \textit{real}
2-vectors in that plane. While we leave the elaboration of this connection
to the interested reader, we feel justif\/ied by it to refer to our f\/indings
as describing ``physical'' (if not necessarily ``realistic'') many-body problems
in the plane.

\subsection{A \textit{solvable} system of two matrix ODEs}\label{section2.1}

In this subsection  we discuss the following system of two matrix ODEs,
satisf\/ied by the two $N\times N$ matrices $U$  and~$V$:
\begin{gather}
\dot{U}=\alpha  U+\beta  U^{2}+\gamma  V+\eta  ( U V+V U )  ,\qquad
\dot{V}=\rho  V .  \label{EqUV}
\end{gather}

\noindent
{\bf Notation.} Upper-case Latin letters generally denote $N\times N$
matrices (unless otherwise indicated), with the (scalar!) $N$ being
throughout an arbitrary \textit{positive integer}. Here and below the
matrices are time-dependent (unless otherwise indicated), in particular $%
U\equiv U(t)$, $V\equiv V(t) $. Lower case letters
generally denote scalars. The $5$ scalar parameters $\alpha $, $\beta $, $%
\gamma $, $\eta $, $\rho $ are time-independent but \textit{a priori}
arbitrary, until specif\/ic restrictions on their values are explicitly
mentioned. Superimposed dots always indicate time dif\/ferentiations.

\begin{remark}
The factor $\beta $ multiplying the term $U^{2}$ in the
right-hand side of the f\/irst of the two matrix ODEs (\ref{EqUV}) could be
eliminated (i.e., replaced by unity) by rescaling $U$\thinspace (and by
correspondingly replacing $\gamma $ with an adjusted value, say $\tilde{\gamma})$; then the factor $\eta $ multiplying the term $ ( U V+V U )$, or the (adjusted) factor $\tilde{\gamma}$ multiplying the
term $V$, could also be eliminated (i.e., replaced by unity) by rescaling~$V $. It is preferable not to do so in order to keep open the possibility to
set one or the other of these parameters, $\eta $~or~$\gamma $, to zero (but
we will never set \textit{both} of them to zero, to avoid decoupling the
time evolution of $U(t) $ from that of~$V(t)$).
Note moreover that the introduction of a constant scalar term (implicitly
multiplied by the $N\times N$ unit matrix~$I$) in the right-hand side of the
f\/irst equation would amount~-- up to a redef\/inition of other parameters~-- to
adding to the matrix $U$ the $N\times N$ unit matrix $I$ multiplied by a~new
(constant) parameter, implying just a constant shift of all the eigenvalues
of the matrix~$U$, a~trivial change not worth pursuing; while the
introduction of two \textit{different} parameters in front of the two terms~$UV$ and $VU$ could be eliminated by adding, in the right-hand side of the
f\/irst of the two matrix ODEs (\ref{EqUV}), the commutator~$[ U,V]
$ of~$U$ and~$V$ times a convenient parameter, without any ef\/fect on the
eigenvalues of~$U$. If some of the parameters in the matrix ODEs (\ref{EqUV}) vanish this model may reduce to one of those that have been previously
treated, see \cite{C2008} (in particular its Section~4.2.2 entitled
``Goldf\/ishing'') and the more recent papers \cite{CC1, CC2, CC3, CC4, CC5, CC6, CC7, CC8, CC9, CC10, CC11, CC12, CC13, CC14, CC15, CC16, CC17}; we do not discard these
models in the following, since we consider in any case worthwhile to exploit
the unif\/ied treatment provided by the present approach based on (\ref{EqUV}). As for the possibility of replacing the right-hand side of the second
matrix ODE (\ref{EqUV}) with a more general function of the matrix $V$, it
is a possibility whose exploration is postponed to a future investigation
(see Section~\ref{section3}).
\end{remark}

\textit{Solution of the system of two matrix ODEs \eqref{EqUV}}. Clearly the
solution of the second of the two matrix ODEs (\ref{EqUV}) reads%
\begin{gather}
V(t) =V_{0} \exp  ( \rho  t )  , \qquad V_{0}\equiv V (0).  \label{SolvV}
\end{gather}

To solve the f\/irst of the two matrix ODEs (\ref{EqUV}) it is convenient to
set
\begin{gather}
U=-\frac{\eta }{\beta } V- ( \beta  F ) ^{-1} \dot{F} ,  \label{UVF}
\end{gather}
obtaining thereby for the $N\times N$ matrix $F\equiv F(t) $ the
following second-order \textit{linear} matrix ODE:
\begin{subequations}
\label{ODEF}
\begin{gather}
\ddot{F}-\alpha  \dot{F}+F W=0 ,  \label{EqF}
\\
W= ( -\alpha  \eta +\beta  \gamma +\eta  \rho  )  V-\eta
^{2} V^{2} .  \label{W}
\end{gather}
\end{subequations}

It is then plain via (\ref{SolvV}) that the \textit{general solution} of
this matrix ODE reads
\begin{gather}
F(t) =F_{+} f_{+} ( t;V_{0} ) +F_{-} f_{-} (t;V_{0} )  ,  \label{Ft}
\end{gather}
where $F_{\pm }$~are two \textit{arbitrary constant} $N\times N$ matrices
and the two \textit{scalar} functions $f_{\pm } ( t;v ) $ (which of
course become $N\times N$ \textit{matrices} when the \textit{scalar} $v$ is
replaced by the $N\times N$ \textit{matrix}~$V_{0}$, see~(\ref{Ft})) are two
independent solutions of the \textit{scalar} second-order \textit{linear} ODE%
\begin{gather}
\ddot{f}-\alpha  \dot{f}+\big[  ( -\alpha  \eta +\beta  \gamma +\eta
 \rho  )  v \exp  ( \rho  t ) -\eta ^{2} v^{2} \exp  (
2 \rho  t ) \big]  f=0 .  \label{ODEf}
\end{gather}
And it is easily seen that two independent solutions of this ODE are given
by the following formulas:
\begin{subequations}
\label{ff}
\begin{gather}
f_{+} ( t;v ) =\exp \left[ -\frac{\eta  v}{\rho } \exp  ( \rho
 t ) \right]  \Phi \left( -\frac{\beta  \gamma }{2 \eta  \rho };1-\frac{\alpha }{\rho };\frac{2\eta v}{\rho }\exp ( \rho t) \right),
\label{f+}
\\
f_{-} ( t;v )  = \exp \left[ -\frac{\eta  v}{\rho } \exp  (
\rho  t ) \right]  \exp  ( \alpha  t )      \Phi \left( \frac{2 \alpha  \eta -\beta  \gamma }{2 \eta  \rho };1+
\frac{\alpha }{\rho };\frac{2 \eta  v}{\rho } \exp  ( \rho  t )
\right)  .
\end{gather}
\end{subequations}
Here the (scalar!) function $\Phi ( a;c;z) $ is the conf\/luent
hypergeometric function (see, for instance,~\cite{HTF1}).

Note that the formula (\ref{UVF}) with (\ref{Ft}) implies the following
\textit{explicit solution} of the \textit{initial-value problem} for the
matrix $U(t) $:
\begin{subequations}
\label{ExplSolUt}
\begin{gather}
 U(t) =-\frac{1}{\beta }\left\{ \eta  V_{0}\exp  ( \rho
 t )   +\big[ f_{+} ( t;V_{0} ) +C~f_{-} ( t;V_{0} ) %
\big] ^{-1} \big[ \dot{f}_{+} ( t;V_{0} ) +C \dot{f}_{-} (
t;V_{0} ) \big] \right\}  ,  \label{Ut}
\end{gather}
with the time-independent matrix $C$ def\/ined in terms of the initial data $U( 0) $ and $V_{0}\equiv V( 0) $ (see (\ref{SolvV}))
as follows:
\begin{gather}
C  = -\big\{ f_{+} ( 0;V_{0} )  [ \beta  U ( 0 )
+\eta  V_{0} ] +\dot{f}_{+} ( 0;V_{0} ) \big\}
  \big\{ f_{-} ( 0;V_{0} )  [ \beta  U ( 0 )
+\eta  V_{0} ] +\dot{f}_{-} ( 0;V_{0} ) \big\} ^{-1}\! .\!\!\!
\end{gather}
\end{subequations}

\begin{remark}\label{remark2.2}
The fact that the expression (\ref{UVF}) (with (\ref{SolvV}) and (\ref{ODEF})) of the matrix $U(t) $ satis\-f\/ies~(\ref{EqUV}), and likewise that the functions $f_{\pm } ( t;v ) $ (see~(\ref{ff})) satisfy the ODE (\ref{ODEf}), can be easily verif\/ied. Of course
these formulas are valid for \textit{generic} values of the parameters that
appear in them, excluding special cases~-- such as vanishing values of $\rho $
or $\eta $ (see~(\ref{ff})) or of $\beta $ (see~(\ref{Ut})), or values of $%
\alpha $ and $\rho $ such that $\alpha /\rho $ is an \textit{integer} (in
which case the two functions $f_{\pm }\left( t;v\right) $ do \textit{not}
provide two \textit{independent} solutions of the ODE~(\ref{ODEf}): see
Section~6.7 of~\cite{HTF1})~-- which are clearly problematic (although the
formulas may in these cases be reinterpreted via appropriate limiting
procedures). We ignore hereafter these issues, even when listing below
\textit{solvable} $N$-body models a few of which belong to these problematic
cases. Indeed in this paper we mainly limit our consideration to the \textit{identification} of \textit{solvable} $N$-body problems; these f\/indings open
the way to obtaining a rather detailed understanding of their actual
behaviors (see for instance the following Remark~\ref{remark2.3}), but such
analyses exceed the scope of this paper: they should be done on a
case-by-case basis, and shall perhaps be postponed to the moment when some
of these $N$-body models evoke a specif\/ic, theoretical or applicative,
interest.
\end{remark}

\begin{remark}\label{remark2.3}
If the parameter $\rho $ is purely \textit{imaginary}
and the parameter $\alpha $ is \textit{real} and \textit{negative},
\begin{subequations}
\begin{gather}
\rho =\frac{2 \pi  \text{\textbf{i}}}{T} , \qquad \alpha <0  \label{rhoT}
\end{gather}%
(of course with $T$ \textit{real} and \textit{nonvanishing,} and \textbf{i}
the \textit{imaginary unit}, \textbf{i}$^{2}=-1$), then clearly the matrix $U(t) $ (see (\ref{ExplSolUt}) with (\ref{ff})) is \textit{asymptotically} \textit{isochronous} (i.e., \textit{asymptotically periodic}
with the period $T$ independent of the initial data): indeed in this case,
as $t\rightarrow +\infty $,
\begin{gather}
U(t) =U_{+}(t) +O [ \exp  ( \alpha  t ) ]
\end{gather}%
with $U_{+}(t) $ given by the formula (\ref{ExplSolUt}) with $%
C=0,$
\begin{gather}
U_{+}(t) =-\frac{1}{\beta }\left\{ \eta  V_{0}\exp  ( \rho
 t ) + [ f_{+} ( t;V_{0} )  ] ^{-1}  [ \dot{f}
_{+} ( t;V_{0} )  ] \right\}  ,
\end{gather}%
hence (see (\ref{rhoT}) and (\ref{f+})) it is \textit{periodic} with period $T$,
\begin{gather}
U_{+} ( t+T ) =U_{+}(t)  .
\end{gather}
\end{subequations}

This observation entails of course the property of \textit{asymptotic
isochrony} for the \textit{solvable} $N$-body models identif\/ied below
featuring parameters $\alpha $ and $\rho $ which satisfy the conditions (\ref%
{rhoT}).
\end{remark}

\subsection{Identif\/ication of \textit{solvable} many-body models}\label{section2.2}

The starting point is to introduce the $N$ eigenvalues $z_{n}(t)
$ of the matrix $U(t) ,$ and the corresponding \textit{%
diagonalizing} matrix $R(t) ,$ by setting
\begin{subequations}
\label{UVM}
\begin{gather}
U(t) =R(t)  Z(t)   [ R (
t )  ] ^{-1} ,  \qquad  Z(t) =\text{diag} [ z_{n} (
t )  ]  .  \label{UZ}
\end{gather}%
Here and hereafter indices such as $n$, $m$, $\ell$, $k$ run over the integers from~$1$ to~$N$ (unless otherwise indicated).

Likewise we set%
\begin{gather}
V(t) =R(t)  Y(t)   [ R (
t )  ] ^{-1} ,  \qquad  Y_{nn}(t) =y_{n}(t)  ,
\label{VY}
\end{gather}%
and we introduce the $N\times N$ matrix $M(t) $ by setting
\begin{gather}
M(t) = [ R(t)  ] ^{-1} \dot{R} (
t )  ,  \qquad  M_{nn}(t) =\mu _{n}(t)  .  \label{M}
\end{gather}
\end{subequations}

\begin{remark}\label{remark2.4}
The diagonalizing matrix $R(t) $ is def\/ined
up to right-multiplication by an \textit{arbitrary diagonal} $N\times N$
matrix $D(t) $, since the replacement of $R(t) $ by $\tilde{R}(t) =R(t)  D(t) $ does not
af\/fect~(\ref{UZ}). But it changes $M(t) $ (see (\ref{M})) into $%
\tilde{M}(t) = [ \tilde{R}(t)  ] ^{-1}
\dot{\tilde{R}}(t) $ implying the following change
of its diagonal elements: $\mu _{n}(t) \Longrightarrow \tilde{\mu }_{n}(t) =\mu _{n}(t) +\dot{d}_{n}(t)
/d_{n}(t) $, where the $N$ quantities~$d_{n}(t) $
are the, \textit{a priori} arbitrary, elements of the diagonal matrix $D(t) $. Hence we retain the privilege to assign at our
convenience (see below) the diagonal elements~$\mu _{n}(t) $ of
the matrix~$M(t) $.
\end{remark}

It is now easily seen that via these assignments (\ref{UVM}) the two matrix
ODEs (\ref{EqUV}) get rewritten as follows:
\begin{gather}
\dot{Z}+ [ M,Z ] =\alpha  Z+\beta  Z^{2}+\gamma  Y+\eta   ( Z
Y+Y Z )  ,  \qquad  \dot{Y}+ [ M,Y ] =\rho  Y .  \label{EqZY}
\end{gather}
Here and hereafter the notation $ [ A,B ] $ denotes the commutator
of the two matrices $A$ and $B$: $ [ A,B ] \equiv A B-B A$.

Let us now look, componentwise, at the \textit{diagonal} and \textit{off-diagonal} elements of these two matrix ODEs.

The \textit{diagonal} part of the f\/irst of the two ODEs (\ref{EqZY}) reads
\begin{subequations}
\begin{gather}
\dot{z}_{n}=\alpha  z_{n}+\beta  z_{n}^{2}+\gamma  y_{n}+2 \eta  z_{n} y_{n} ,  \label{zndot}
\end{gather}
implying
\begin{gather}
y_{n}=\frac{\dot{z}_{n}-\alpha  z_{n}-\beta  z_{n}^{2}}{\gamma +2 \eta  z_{n}
} .  \label{yn}
\end{gather}
\end{subequations}

The \textit{off-diagonal} part of the f\/irst of the two ODEs (\ref{EqZY})
reads
\begin{subequations}
\begin{gather}
-\left( z_{n}-z_{m}\right)  M_{nm}= [ \gamma +\eta   (
z_{n}+z_{m} )  ]  Y_{nm} ,  \qquad  n\neq m ,
\end{gather}%
implying%
\begin{gather}
M_{nm}=-\left[ \frac{\gamma +\eta  \left( z_{n}+z_{m}\right) }{z_{n}-z_{m}}%
\right]  Y_{nm} ,  \qquad  n\neq m .  \label{Mnm}
\end{gather}

The \textit{diagonal} part of the second of the two ODEs (\ref{EqZY}) reads
\end{subequations}
\begin{subequations}
\begin{gather}
\dot{y}_{n}=\rho  y_{n}+\sum_{\ell =1, \ell \neq n}^{N}\left( Y_{n\ell
} M_{\ell n}-M_{n\ell } Y_{\ell n}\right)  ,
\end{gather}%
implying, via (\ref{Mnm}),
\begin{gather}
\dot{y}_{n}=\rho  y_{n}+2 \sum_{\ell =1, \ell \neq n}^{N}\left\{ Y_{n\ell
} Y_{\ell n} \left[ \frac{\gamma +\eta  ( z_{n}+z_{\ell }) }{z_{n}-z_{\ell }}\right] \right\}  .  \label{yndot}
\end{gather}
\end{subequations}

Hence, by time-dif\/ferentiation of (\ref{zndot}) we see via (\ref{yn}) and (\ref{yndot}) that the coordina\-tes~$z_{n}(t) $ satisfy the
following system of~$N$ equations of motion of \textit{Newtonian} type
(``acceleration equal force'', with one-body and two-body forces):
\begin{gather}
 \ddot{z}_{n}=-\alpha  \rho  z_{n}-\beta  \rho  z_{n}^{2}+ ( \alpha
+\rho  )  \dot{z}_{n}+2 \beta  \dot{z}_{n} z_{n}
 +\frac{2 \eta  \dot{z}_{n} \left( \dot{z}_{n}-\alpha  z_{n}-\beta
 z_{n}^{2}\right) }{\gamma +2 \eta  z_{n}}  \nonumber \\
\hphantom{\ddot{z}_{n}=}{}
+2  ( \gamma +2 \eta  z_{n} )  \sum_{\ell =1, \ell \neq
n}^{N}\left\{ Y_{n\ell } Y_{\ell n} \left[ \frac{\gamma +\eta   (
z_{n}+z_{\ell } ) }{z_{n}-z_{\ell }}\right] \right\}  .
\label{zndotdot}
\end{gather}

But in these equations of motion the role of ``two-body coupling constants''
is played by the quantities $Y_{n\ell } Y_{\ell n}$ (with $\ell \neq n$)
which are in fact \textit{time-dependent}. Indeed the time evolution of the
\textit{off-diagonal} elements $Y_{nm}$ of the matrix $Y$ is determined by
the \textit{off-diagonal} part of the second of the two ODEs~(\ref{EqZY})
which componentwise read
\begin{subequations}
\begin{gather}
\dot{Y}_{nm}=\rho  Y_{nm}+\sum_{k=1}^{N}\left(
Y_{nk} M_{km}-M_{nk} Y_{km}\right)  ,  \qquad  n\neq m ,
\end{gather}%
yielding, via (\ref{VY}), (\ref{M}), (\ref{yn}), (\ref{Mnm}) and a bit of
algebra,
\begin{gather}
\frac{\dot{Y}_{nm}}{Y_{nm}}=\alpha +\rho +\beta   ( z_{n}+z_{m} ) -
\frac{\dot{z}_{n}-\dot{z}_{m}}{z_{n}-z_{m}}
+\frac{\eta  \left( \dot{z}_{n}-\alpha  z_{n}-\beta  z_{n}^{2}\right) }{
\gamma +2 \eta  z_{n}}+\frac{\eta  \left( \dot{z}_{m}-\alpha  z_{m}-\beta
 z_{m}^{2}\right) }{\gamma +2 \eta  z_{m}}  \nonumber \\
\hphantom{\frac{\dot{Y}_{nm}}{Y_{nm}}=}{}
-\mu _{n}+\mu _{m} +\sum_{\ell =1, \ell \neq n,m}^{N}\left\{ \frac{Y_{n\ell } Y_{\ell m}}{Y_{nm}%
} \left[ 2 \eta + ( \gamma +2 \eta  z_{\ell } ) \left( \frac{1}{z_{n}-z_{\ell }}+\frac{1}{z_{m}-z_{\ell }}\right) \right] \right\}  , \nonumber\\
n\neq m .    \label{Ynmdot}
\end{gather}
\end{subequations}

So, in order that (\ref{zndotdot}) become the $N$ \textit{Newtonian}
equations of motion of a genuine $N$-body problem one must either provide
some ``physical interpretation'' for the quantities $Y_{n\ell }$ (with $\ell
\neq n$)~-- possibly in terms of internal degrees of freedom, an alternative
we do not pursue in this paper~-- or f\/ind a way to ``get rid'' of these
quantities, i.e.\ f\/ind a way to express them~-- if at all possible~-- via the $N$ coordinates $z_{m}\equiv z_{m}(t) $ and possibly also the $N$
velocities $\dot{z}_{m}\equiv \dot{z}_{m}(t) $. Previous
experience~\cite{C2001, C2008, CC1, CC2, CC3, CC4, CC5, CC6, CC7, CC8, CC9,
CC10, CC11, CC12, CC13, CC14, CC15, CC16, CC17} suggest that two types of
\textit{ans\"{a}tze} are the appropriate starting points to try and achieve
this goal.

\subsubsection{First ansatz}

The f\/irst \textit{ansatz} reads
\begin{gather}
Y_{nm}=\frac{( g_{n} g_{m}) ^{1/2}}{z_{n}-z_{m}} ,\qquad   n\neq m ,
\label{Ans1}
\end{gather}
where we reserve the option to make a convenient assignment for the $N$
functions $g_{n}$ of the coordinate $z_{n}$, $g_{n}\equiv g_{n} (
z_{n} ) \equiv g_{n} [ z_{n}(t)  ] $.

The insertion of this \textit{ansatz} in (\ref{Ynmdot}) yields, after a bit
of trivial algebra,
\begin{gather}
\frac{\dot{g}_{n}}{2 g_{n}}-\frac{\alpha +\rho }{2}-\beta  z_{n}-\frac{\eta
 \left( \dot{z}_{n}-\alpha  z_{n}-\beta  z_{n}^{2}\right) }{\gamma +2 \eta
 z_{n}}+ (  ( n\rightarrow m )  )    \nonumber \\
\qquad {} =\frac{g_{n}  ( \gamma +2 \eta  z_{m} ) -g_{m}  ( \gamma
+2 \eta  z_{n} ) }{( z_{n}-z_{m}) ^{2}}-\mu _{n}+\mu _{m}
\nonumber \\
\qquad\quad{} +\sum_{\ell =1, \ell \neq n}^{N}\left\{ g_{\ell } \left[ \frac{\gamma
+2 \eta  z_{n}}{( z_{n}-z_{\ell }) ^{2}}\right] \right\}
-\sum_{\ell =1, \ell \neq m}^{N}\left\{ g_{\ell } \left[ \frac{\gamma
+2 \eta  z_{m}}{( z_{m}-z_{\ell }) ^{2}}\right] \right\}
 ,  \qquad n\neq m .   \label{EqAns1}
\end{gather}
Here and throughout the notation $+ (  ( n\rightarrow m )
 ) $ indicates the addition of whatever comes before it, with the index~$n$ replaced by~$m$.

We now take advantage of the freedom (see Remark~\ref{remark2.4}) to assign
the diagonal elements $\mu _{n}$ of the matrix $M$ by setting
\begin{gather}
\mu _{n}=\sum_{\ell =1, \ell \neq n}^{N}\left\{ g_{\ell } \left[ \frac{\gamma +2 \eta  z_{n}}{ ( z_{n}-z_{\ell } ) ^{2}}\right] \right\}  ,
\label{mun}
\end{gather}
and we moreover make the assignment%
\begin{gather}
g_{n}=g  ( \gamma +2 \eta  z_{n} )  ,  \label{gn}
\end{gather}%
with $g$ an arbitrary constant (i.e., $\dot{g}=0$). Thereby the system of $N ( N-1 ) $ equations (\ref{EqAns1}) gets reduced to the following,
much simpler, system of only $N$ algebraic equations:
\begin{gather}
-\frac{\alpha +\rho }{2}-\beta  z_{n}+\frac{\eta  \left( \alpha  z_{n}+\beta
 z_{n}^{2}\right) }{\gamma +2 \eta  z_{n}}=0 ,
\end{gather}%
which amounts to the following $3$ equations (recall that we exclude the
uninteresting possibility that $\gamma $ and $\eta $ both vanish):
\begin{subequations}
\begin{gather}
\beta  \eta =0 ,
\\
\beta  \gamma +\eta  \rho =0 ,
\\
\gamma  \left( \alpha +\rho \right) =0 .
\end{gather}
\end{subequations}
This boils down to the following $3$ possibilities:
\begin{subequations}
\begin{gather}
\alpha =\beta =\rho =0 ,
\end{gather}
or
\begin{gather}
\beta =\gamma =\rho =0 ,
\end{gather}
or%
\begin{gather}
\beta =\eta =0 ,  \qquad  \rho =-\alpha  .
\end{gather}
\end{subequations}
The corresponding \textit{solvable} $N$-body models~-- obtained by inserting (%
\ref{Ans1}) with (\ref{gn}) and with these assignments of the parameters in
the \textit{Newtonian} equations of motion (\ref{zndotdot})~-- read, in the
f\/irst~2 of these~3 cases~-- after conveniently setting
\begin{subequations}
\begin{gather}
\gamma +2 \eta  z_{n}(t) =\exp  [ 2 c \zeta _{n} (
t )  ]
\end{gather}
with $c$ an arbitrary nonvanishing constant~-- as follows:
\begin{gather}
\ddot{\zeta}_{n}=\left( \frac{g^{2} \eta ^{4}}{c^{2}}\right)  \frac{d}{d \zeta _{n}}\sum_{\ell =1, \ell \neq n}^{N}\sinh ^{-2} [ c ( \zeta
_{n}-\zeta _{\ell } )  ]  ;
\end{gather}
\end{subequations}
while in the third of these 3 cases they read
\begin{gather}
\ddot{z}_{n}=\alpha ^{2} z_{n}+g^{2} \gamma ^{4} \frac{d}{d z_{n}}\sum_{\ell
=1, \ell \neq n}^{N} ( z_{n}-z_{\ell } ) ^{-2} .
\end{gather}
But these are well-known \textit{solvable} $N$-body problems, see for
instance \cite{C2001}. Hence we conclude that from the matrix system (\ref%
{EqUV}) \textit{no} \textit{new} \textit{solvable} many-body models are
obtained via the \textit{ansatz} (\ref{Ans1}).

\subsubsection{Second ansatz}

We proceed then to consider a second \textit{ansatz}, reading
\begin{gather}
Y_{nm}=\left[ g_{n} g_{m} \left( \dot{z}_{n}+f_{n}\right)  \left( \dot{z}%
_{m}+f_{m}\right) \right] ^{1/2} ,  \qquad n\neq m ,  \label{Ans2}
\end{gather}%
with $g_{n}$ and $f_{n}$ functions of the coordinate $z_{n}$ that we reserve
to assign later: $g_{n}\equiv g_{n} ( z_{n} ) \equiv g_{n} [
z_{n}(t)  ] $, $f_{n}\equiv f_{n} ( z_{n} ) \equiv
f_{n} [ z_{n}(t)  ] $. Its insertion in (\ref{Ynmdot})
yields, again after a bit of trivial algebra and now the assignment $\mu
_{n}=0$ (again justif\/ied by  Remark~\ref{remark2.4}), the following system of $N ( N-1 ) $ second-order ODEs:%
\begin{gather}
\frac{1}{2} \left( \frac{\ddot{z}_{n}+\dot{f}_{n}}{\dot{z}_{n}+f_{n}}+\frac{
\dot{g}_{n}}{g_{n}}-\alpha -\rho \right) -\beta  z_{n}-\eta  \frac{\dot{z}
_{n}-\alpha  z_{n}-\beta  z_{n}^{2}}{\gamma +2 \eta  z_{n}}   \nonumber \\
\qquad{} +\eta   ( \dot{z}_{n}+f_{n} )  g_{n}-\sum_{\ell =1, \ell \neq
n}^{N}\left\{  ( \dot{z}_{\ell }+f_{\ell } )  g_{\ell } \left[
\frac{\gamma +\eta   ( z_{n}+z_{\ell } ) }{z_{n}-z_{\ell }}\right]
\right\}    \nonumber \\
\qquad{} -\frac{\dot{z}_{n}  [ g_{n}  ( \gamma +2 \eta  z_{n} ) -1 ]
+f_{n} g_{n}  ( \gamma +2 \eta  z_{n} ) }{z_{n}-z_{m}}+ (
 ( n\rightarrow m )  ) =0 , \qquad
n\neq m .    \label{EqAns2}
\end{gather}

To reduce this system we clearly must now set
\begin{subequations}
\label{fg}
\begin{gather}
g_{n}=\frac{1}{\gamma +2 \eta  z_{n}} ,  \label{g}
\\
f_{n}=f^{( 0) }+f^{( 1) } z_{n}+f^{( 2)
} z_{n}^{2} ,  \label{fn}
\end{gather}
\end{subequations}
where $f^{( 0) }$, $f^{( 1) }$, $f^{( 2) }$ are
$3$ \textit{constant} parameters that we reserve to assign later. Thereby
the system of $N ( N-1 ) $ ODEs (\ref{EqAns2}) gets transformed
into the following system of (only) $N$ ODEs:
\begin{gather}
 \ddot{z}_{n}-2  ( \dot{z}_{n}+f_{n} )  \sum_{\ell =1, \ell \neq
n}^{N}\left\{ \frac{ ( \dot{z}_{\ell }+f_{\ell } )   [ \gamma
+\eta   ( z_{n}+z_{\ell } )  ] }{( z_{n}-z_{\ell } )
  ( \gamma +2 \eta  z_{\ell } ) }\right\}  \nonumber \\
\qquad{} = \left[ \alpha +\rho +2 \beta  z_{n}+\frac{2 \eta  \left( \dot{z}
_{n}-\alpha  z_{n}-\beta  z_{n}^{2}-f_{n}\right) }{\gamma +2 \eta  z_{n}}
\right]   ( \dot{z}_{n}+f_{n} )
 +\big( f^{( 1) }+2 f^{( 2) } z_{n}\big)  f_{n} .\!\!\!
\label{EqAns2bis}
\end{gather}
Note that to write this system in more compact form we employed a mixed
notation, using sometimes the functions $f_{n}$ or $f_{\ell }$ instead of
their explicit expressions, see~(\ref{fn}).

To complete the task of ascertaining for which values of the (so far
arbitrary) $8$ \textit{constant} parameters $\alpha $, $\beta $, $\gamma $, $%
\eta $, $\rho $, $f^{( 0) }$, $f^{( 1) }$, $f^{(
2) }$ the system of $N( N-1) $ ODEs (\ref{EqAns2bis}) is
satisf\/ied we must utilize the equations of motion (\ref{zndotdot}) which,
via the \textit{ansatz} (\ref{Ans2}) with (\ref{fg}), now read
\begin{gather}
 \ddot{z}_{n}-2  ( \dot{z}_{n}+f_{n} )  \sum_{\ell =1, \ell \neq
n}^{N}\left\{ \frac{( \dot{z}_{\ell }+f_{\ell })  [ \gamma
+\eta  ( z_{n}+z_{\ell }) ] }{( z_{n}-z_{\ell })
( \gamma +2 \eta  z_{\ell }) }\right\}  \nonumber \\
\qquad{} = -\alpha  \rho  z_{n}-\beta  \rho  z_{n}^{2}+ ( \alpha +\rho  )
\dot{z}_{n}+2 \beta  \dot{z}_{n} z_{n}
 +\frac{2 \eta  \dot{z}_{n} \left( \dot{z}_{n}-\alpha  z_{n}-\beta
 z_{n}^{2}\right) }{\gamma +2 \eta  z_{n}} .  \label{EqMotGold1}
\end{gather}
Comparison of this system of $N$ ODEs to the system (\ref{EqAns2bis}) yields
the following system of $N$ algebraic equations (note that~-- as it were,
``miraculously''~-- the velocities $\dot{z}_{n}$ have disappeared from these
equations):
\begin{gather}
\alpha  \rho  z_{n}+\beta  \rho  z_{n}^{2}+\big[ \alpha +\rho +f^{(
1) }+2 \big( \beta +f^{( 2) }\big)  z_{n}\big]  f_{n}
=\frac{2 \eta  \left( \alpha  z_{n}+\beta  z_{n}^{2}+f_{n}\right)  f_{n}}{\gamma +2 \eta  z_{n}} .   \label{Cond2}
\end{gather}

It is now clear that, in order to satisfy this system~-- identically, i.e.\
for any values of the coordinates $z_{n}$~-- one must set either
\begin{subequations}
\label{case(i)}
\begin{gather}
\text{case~$(i)$}:   \ f_{n}= ( a+b z_{n} )   ( \gamma
+2 \eta  z_{n} )  \label{fab1}
\end{gather}
implying (see (\ref{fn}))
\begin{gather}
\text{case $(i)$}: \ f^{( 0) }=a \gamma  ,  \qquad  f^{(1) }=2 a \eta +b \gamma  , \qquad  f^{( 2) }=2 b \eta  ,
\end{gather}
\end{subequations}
or
\begin{subequations}
\label{case(ii)}
\begin{gather}
\text{case $(ii)$}:  \ f_{n}=-\alpha  z_{n}-\beta  z_{n}^{2}+ (
a+b z_{n} )   ( \gamma +2 \eta  z_{n} )  \label{fab2}
\end{gather}
implying (see (\ref{fn}))
\begin{gather}
\text{case $(ii)$}: \ f^{( 0) }=a \gamma  ,  \qquad f^{(1) }=2 a \eta +b \gamma -\alpha  , \qquad  f^{( 2) }=2 b \eta
-\beta  .
\end{gather}
\end{subequations}

So, in both cases, we hereafter only retain the freedom to assign the $2$
constants $a$, $b$ rather than the $3$ constants $f^{( 0) }$, $ f^{( 1) }$, $ f^{( 2) }$.

Clearly in  case $(i)$ we get the following system of $N$ algebraic
equations:
\begin{subequations}
\begin{gather}
\alpha  \rho  z_{n}+\beta  \rho  z_{n}^{2}+ [ \alpha +\rho +2 a \eta
+b \gamma +2 ( \beta +2 b \eta  )  z_{n} ]
  \left[ a \gamma + ( 2 a \eta +b \gamma  )  z_{n}+2 b \eta
 z_{n}^{2}\right]    \nonumber \\
\qquad =2 \eta  \left[ \alpha  z_{n}+\beta  z_{n}^{2}+ ( a+b z_{n} )
 ( \gamma +2 \eta  z_{n} ) \right]   ( a+b z_{n} )  ,
\end{gather}
and likewise in  case $(ii)$
\begin{gather}
\alpha  \rho  z_{n}+\beta  \rho  z_{n}^{2}+ [ \rho +2 a \eta +b \gamma
+4 b \eta  z_{n} ]
\left[ a \gamma + ( 2 a \eta +b \gamma -\alpha  )  z_{n}+ (
2 b \eta -\beta  )  z_{n}^{2}\right]    \nonumber \\
\qquad =2 \eta  \left[ -\alpha  z_{n}-\beta  z_{n}^{2}+ ( a+b z_{n} )
 ( \gamma +2 \eta  z_{n} ) \right]   ( a+b z_{n} )  .
\end{gather}
\end{subequations}
Hence in  case $(i)$ the following set of $4$ nonlinear algebraic
equations must be satisf\/ied by the $7$~parameters $a$, $b$, $\alpha$, $\beta$, $\gamma$, $\eta $, $\rho $:
\begin{subequations}
\label{case1}
\begin{gather}
b \eta   ( \beta +2 b \eta  ) =0 ,
\\
 ( \beta +2 b \eta  )   ( \rho +2 a \eta +2 b \gamma  )
=0 ,
\\
 ( \alpha +2 a \eta +b \gamma  )   ( \rho +b \gamma  )
+2 a  ( \beta +b \eta  )  \gamma =0 ,
\\
a  ( \alpha +\rho +b \gamma  )  \gamma =0 .
\end{gather}
\end{subequations}
Likewise in  case $(ii)$ the following set of $4$ nonlinear algebraic
equations must be satisf\/ied by the $7$~parameters $a$, $b$, $\alpha$, $\beta$, $\gamma$, $\eta $, $\rho $:{\samepage
\begin{subequations}
\label{case2}
\begin{gather}
b \eta   ( \beta -2 b \eta  ) =0 ,
\\
b  [ 2 \eta   ( 2 a \eta +2 b \gamma -\alpha +\rho  ) -\beta
 \gamma  ] =0 ,
\\
2 a \eta  \rho +b \gamma   ( 4 a \eta +b \gamma -\alpha +\rho  )
=0 ,
\\
a  ( \rho +b \gamma  )  \gamma =0 .
\end{gather}
\end{subequations}

}

Then a trivial if rather tedious computation yields, in  case $(i)$
respectively in  case $(ii)$, the solutions reported in  Table~2.1 respectively in  Table~2.2 (note that for $\alpha =\beta =\eta
=0$ these two cases coincide, so the corresponding results are only included
in  Table~2.2).

\begin{figure}[t!]{\centering
{\bf Table 2.1}\label{table2.1}

\vspace{1mm}

$\left\vert
\begin{array}{cccccccc}
\# & a & b & \alpha & \beta & \gamma & \eta & \rho \\
1 & 0 & 0 & \ast & \ast & \ast & \ast & 0 \\
2 & \ast & 0 & \ast & 0 & 0 & \ast & 0 \\
3 & \ast & 0 & -2 a \eta & 0 & 0 & \ast & \ast \\
4 & \ast & 0 & -2 a \eta & 0 & \ast & \ast & 2 a \eta \\
5 & \ast & 0 & -2 a \eta & \ast & 0 & \ast & -2 a \eta \\
6 & \ast & 0 & 2 a \eta & 4 a \eta ^{2}/\gamma & \ast & \ast & -2 a \eta \\
7 & 0 & \ast & -b \gamma & 0 & \ast & 0 & \ast \\
8 & \ast & \ast & -b \gamma & 0 & \ast & 0 & 0 \\
9 & 0 & \ast & \ast & 0 & \ast & 0 & -b \gamma \\
10 & 0 & \ast & -b \gamma & \ast & \ast & 0 & -2 b \gamma \\
11 & \ast & \ast & b \gamma & b^{2}\gamma /a & \ast & 0 & -2 b \gamma \\
12 & 0 & \ast & -b \gamma & -2 b \eta & \ast & \ast & \ast \\
13 & 0 & \ast & \ast & -2 b \eta & \ast & \ast & -b \gamma \\
14 & \ast & \ast & \ast & -2 b \eta & 0 & \ast & 0 \\
15 & \ast & \ast & -2 a \eta & -2 b \eta & 0 & \ast & \ast \\
16 & -b \gamma & \ast & \ast & -2 b \eta & \ast & \ast & -\alpha -b \gamma
\\
17 & \ast & \ast & -2 a \eta & -2 b \eta & \ast & \ast & -\alpha -b \gamma%
\end{array}%
\right\vert $

}

\vspace{1mm}

{\small \noindent
This table 
indicates the $17$
sets of values to be assigned to the $7$ parameters $a$, $b$, $\alpha$, $\beta$, $\gamma$, $\eta $, $\rho $ in order to satisfy the $4$ equations
characterizing case~$(i)$, see~(\ref{case1}). Cases with $\gamma
=\eta =0$ are excluded. Asterisks indicate that the corresponding parameters
can be assigned freely. Note that each of the $7$ lines $1$, $12$, $13$, $14 $, $15$, $16$, $17$ assigns values to only $3$ of the $7$ parameters,
while each of the other $10$ lines assigns values to $4$ of the $7$
parameters.}
\end{figure}

\begin{figure}[t!]{\centering

{\bf Table 2.2} \label{table2.2}

\vspace{1mm}

$\left\vert
\begin{array}{cccccccc}
\# & a & b & \alpha & \beta & \gamma & \eta & \rho \\
1 & 0 & 0 & \ast & \ast & \ast & \ast & \ast \\
2 & \ast & 0 & \ast & \ast & \ast & \ast & 0 \\
3 & 0 & \ast & b \gamma +\rho & 0 & \ast & 0 & \ast \\
4 & \ast & \ast & 0 & 0 & \ast & 0 & -b \gamma \\
5 & 0 & \ast & b \gamma +\rho & 2 b \eta & \ast & \ast & \ast \\
6 & 0 & \ast & \rho & 2 b \eta & 0 & \ast & \ast \\
7 & \ast & \ast & 2 a \eta & 2 b \eta & \ast & \ast & -b \gamma%
\end{array}%
\right\vert $

}

\vspace{1mm}

{\small\noindent
This table 
indicates the $7$
sets of values to be assigned to the $7$ parameters $a$, $b$, $\alpha$, $\beta$, $\gamma$, $\eta $, $\rho $ in order to satisfy the $4$ equations
characterizing case~$(ii)$, see~(\ref{case2}). Cases with $\gamma
=\eta =0$ are excluded. Asterisks indicate that the corresponding parameters
can be assigned freely. Note that each of the lines $1$ and $2$ assigns
values to only $2$ of the $7$ parameters, lines $3$, $4$ and $6$ assign
values to $4$ of the $7$ parameters, and lines $5$ and $7$ assign values to $%
3$ of the $7$ parameters.}

\end{figure}

We therefore conclude that the many-body models of goldf\/ish type
characterized by the \textit{Newtonian} equations of motion (\ref{EqMotGold1}) (or, equivalently, (\ref{EqAns2bis})) are \textit{solvable} provided
either the quantities $f_{n}$ are expressed by (\ref{fab1}) and the $7$
parameters $a$, $b$, $\alpha$, $\beta$, $\gamma$, $\eta $, $\rho $ are
consistent with Table~2.1 or the quantities $f_{n}$ are expressed
by (\ref{fab2}) and the $7$ parameters $a$, $b$, $\alpha$, $\beta$, $\gamma$, $\eta $, $\rho $ are consistent with Table~2.2. There are
therefore altogether $24$ \textit{solvable} models. Some of these models are
however \textit{not new} (in particular, when some parameters vanish);
moreover in some cases the solution (\ref{ExplSolUt}) with (\ref{ff}) of the
matrix equation (\ref{EqUV}) is only valid in a limiting sense (see Remark~\ref{remark2.2}). We leave these issues to whoever will be interested~-- possibly
in specif\/ic, theoretical or applicative, contexts~-- in more detailed
investigations of anyone of these models; our focus in this paper is rather
in the unif\/ied treatment, and the simultaneous display (see the appendices
below), of all of them (except for the elimination, as already mentioned, of
the cases with $\gamma =\eta =0$).

It is convenient to write the corresponding equations of motion in two
dif\/ferent ways, depending whether the parameter $\eta $ does or does not
vanish.

If the parameter $\eta $ vanishes, $\eta =0$, the equations of motion read
as follows:
\begin{gather}
\ddot{z}_{n}  = -\alpha  \rho  z_{n}-\beta  \rho  z_{n}^{2}+( \alpha
+\rho )  \dot{z}_{n}+2 \beta  \dot{z}_{n} z_{n}   +2 ( \dot{z}_{n}+f_{n})  \sum_{\ell =1, \ell \neq n}^{N}\left[
\frac{\left( \dot{z}_{\ell }+f_{\ell }\right) }{( z_{n}-z_{\ell
}) }\right]  \label{EqMotznEtaEqZero}
\end{gather}
with the following assignments corresponding respectively to  case $(i)$ and case~$(ii)$.

In case $(i)$
\begin{gather}
f_{n}=a \gamma +b \gamma  z_{n}  \label{fnEtaEqZeroCase(i)}
\end{gather}%
and the $7$ parameters $a$, $b$, $\alpha$, $\beta$, $\gamma$, $\eta$, $\rho $
are restricted according to 
Table~2.3 (being the
relevant subcase of  Table~2.1).

\begin{figure}[t!]{\centering
{\bf Table 2.3}\label{table2.3}
\vspace{1mm}

$\left\vert
\begin{array}{cccccccc}
\# & a & b & \alpha & \beta & \gamma & \eta & \rho \\
1 & 0 & \ast & -b \gamma & 0 & \ast & 0 & \ast \\
2 & \ast & \ast & -b \gamma & 0 & \ast & 0 & 0 \\
3 & 0 & \ast & \ast & 0 & \ast & 0 & -b \gamma \\
4 & 0 & \ast & -b \gamma & \ast & \ast & 0 & -2 b \gamma \\
5 & \ast & \ast & b \gamma & b^{2}\gamma /a & \ast & 0 & -2 b \gamma%
\end{array}%
\right\vert $

}

\vspace{1mm}

{\small\noindent
This table 
indicates the $5$
sets of values to be assigned to the $7$ parameters $a$, $b$, $\alpha$, $\beta$, $\gamma$, $\eta$, $\rho $ in (\ref{EqMotznEtaEqZero}) with~(\ref{fnEtaEqZeroCase(i)}). Asterisks indicate that the corresponding parameters
can be assigned freely. Note that in every case there are $3$ free
parameters.}
\end{figure}

In case $(ii)$
\begin{gather}
f_{n}=a \gamma +\left( b \gamma -\alpha \right)  z_{n}-\beta  z_{n}^{2}
\label{fnEtaEqZeroCase(ii)}
\end{gather}%
and the $6$ parameters $a$, $b$, $\alpha$, $\beta$, $\gamma$, $\rho $ are
restricted according to 
Table~2.4 (being the relevant subcase of Table~2.2).

\begin{figure}[t!]{\centering
{\bf Table 2.4}\label{table2.4}

\vspace{1mm}

$\left\vert
\begin{array}{cccccccc}
\# & a & b & \alpha & \beta & \gamma & \eta & \rho \\
1 & 0 & \ast & b \gamma +\rho & 0 & \ast & 0 & \ast \\
2 & \ast & \ast & 0 & 0 & \ast & 0 & -b \gamma%
\end{array}%
\right\vert $

}

\vspace{1mm}


{\small\noindent
This 
table indicates the set of
values to be assigned to the $7$ parameters $a$, $b$, $\alpha$, $\beta$, $\gamma$, $\eta$, $\rho $ in~(\ref{EqMotznEtaEqZero}) with~(\ref{fnEtaEqZeroCase(ii)}). Asterisks indicate that the corresponding parameters
can be assigned freely: in the second line~$a$,~$b$, and $\gamma $ are $3$ free
parameters, in the f\/irst line $b$, $\gamma $ and $\rho $ are $3$ free
parameters.}
\end{figure}

If the parameter $\eta $ does \textit{not} vanish, $\eta \neq 0$, the
equations of motion are more conveniently written in terms of the dependent
variables
\begin{gather}
x_{n}(t) \equiv z_{n}(t) +\frac{\gamma }{2 \eta } ,
\end{gather}
reading then as follows:
\begin{subequations}
\label{EqMotznEtaNotZero}
\begin{gather}
\ddot{x}_{n}  = \frac{\dot{x}_{n}^{2}}{x_{n}}+\lambda  \frac{\dot{x}_{n}}{%
x_{n}}+\beta  \dot{x}_{n} x_{n}+\rho  \left( \dot{x}_{n}+\lambda -\mu
 x_{n}-\beta  x_{n}^{2}\right)  \nonumber \\
 \hphantom{\ddot{x}_{n}  =}{}
 + ( \dot{x}_{n}+f_{n} )  \sum_{\ell =1, \ell \neq n}^{N}\left[
\frac{ ( \dot{x}_{\ell }+f_{\ell } )   ( x_{n}+x_{\ell } )
}{( x_{n}-x_{\ell })  x_{\ell }}\right]
\end{gather}
or equivalently
\begin{gather}
 \ddot{x}_{n}=\big( \lambda -2 f^{(0) }\big)  \frac{\dot{x}_{n}}{x_{n}}+[ ( N-2)  f^{(1) }+\rho ]
 \dot{x}_{n}+\big( -2 f^{(2) }+\beta \big)  \dot{x}_{n} x_{n}
 -\frac{\big( f^{(0) }\big) ^{2}}{x_{n}}+\rho  \lambda\nonumber\\
 \hphantom{\ddot{x}_{n}=}{}
+ ( N-2 )  f^{(0) } f^{(1) }
 +\big[ {-}\rho  \mu + ( N-1 )  \big( f^{(1) }\big)
^{2}-2 f^{(0) } f^{(2) }\big]  x_{n}  \nonumber \\
\hphantom{\ddot{x}_{n}=}{}
 +\big[ {-}\beta  \rho + ( N-2 )  f^{(1) } f^{(2) }\big]  x_{n}^{2}-\big( f^{(2) }\big)
^{2} x_{n}^{3}
 + ( \dot{x}_{n}+f_{n} )  \sum_{k=1}^{N}\left[ \frac{\big( \dot{x}_{k}+f^{(0) }\big) }{x_{k}}+f^{(2) } x_{k}\right]
\nonumber \\
\hphantom{\ddot{x}_{n}=}{}
 +2 \sum_{\ell =1, \ell \neq n}^{N}\left[ \frac{( \dot{x}_{n}+f_{n})  ( \dot{x}_{\ell }+f_{\ell }) }{(
x_{n}-x_{\ell }) }\right]
\end{gather}
with
\begin{gather}
\lambda =\frac{( 2 \alpha  \eta -\beta  \gamma )  \gamma }{4 \eta
^{2}} ,   \qquad \mu =\alpha -\frac{\beta  \gamma }{\eta }  \qquad  \text{so that} \qquad \mu
^{2}=\alpha ^{2}-\frac{\lambda }{4} ,  \label{landamu}
\end{gather}
\end{subequations}
and with the following assignments corresponding respectively to case~$(i)$ and  case $(ii)$.

In case~$(i)$
\begin{gather}
f_{n}=( 2 a \eta -b \gamma )  x_{n}+2 b \eta  x_{n}^{2}
\label{fnEtaEqNotZeroCase(i)}
\end{gather}%
and the $7$ parameters $a$, $b$, $\alpha$, $\beta$,  $\gamma$, $\eta $, $\rho $
are restricted according to 
Table~2.5 (being the
relevant subcase of  Table~2.1).

\begin{figure}[t!]{\centering
{\bf Table 2.5}\label{table2.5}

\vspace{1mm}

$\left\vert
\begin{array}{cccccccc}
\# & a & b & \alpha & \beta & \gamma & \eta & \rho \\
1 & 0 & 0 & \ast & \ast & \ast & \ast & 0 \\
2 & \ast & 0 & \ast & 0 & 0 & \ast & 0 \\
3 & \ast & 0 & -2 a \eta & 0 & 0 & \ast & \ast \\
4 & \ast & 0 & -2 a \eta & 0 & \ast & \ast & 2 a \eta \\
5 & \ast & 0 & -2 a \eta & \ast & 0 & \ast & -2 a \eta \\
6 & \ast & 0 & 2 a \eta & 4 a \eta ^{2}/\gamma & \ast & \ast & -2 a \eta \\
7 & 0 & \ast & -b \gamma & -2 b \eta & \ast & \ast & \ast \\
8 & 0 & \ast & \ast & -2 b \eta & \ast & \ast & -b \gamma \\
9 & \ast & \ast & \ast & -2 b \eta & 0 & \ast & 0 \\
10 & \ast & \ast & -2 a \eta & -2 b \eta & 0 & \ast & \ast \\
11 & -b \gamma & \ast & \ast & -2 b \eta & \ast & \ast & -\alpha -b \gamma
\\
12 & \ast & \ast & -2 a \eta & -2 b \eta & \ast & \ast & -\alpha -b \gamma%
\end{array}%
\right\vert $

}

\vspace{1mm}


{\small \noindent
This table 
indicates the $12$
sets of values to be assigned to the $7$ parameters $a$, $b$, $\alpha$, $\beta$, $\gamma$, $\eta $, $\rho $ in~(\ref{EqMotznEtaNotZero}) with~(\ref{fnEtaEqNotZeroCase(i)}). Asterisks indicate that the corresponding
parameters can be assigned freely (with $\eta \neq 0$). Note that each of
the $7$ lines $1$, 7, 8, 9, 10, 11, 12 assigns values to only $3$ of the $7 $ parameters, while the other $5$ assign values to $4$ of the $7$
parameters.}
\end{figure}

In case $(ii)$
\begin{gather}
f_{n}=\frac{( 2 \alpha  \eta -\beta  \gamma )  \gamma }{4 \eta
^{2}}+\left( 2 a \eta -b \gamma -\alpha +\frac{\beta  \gamma }{\eta }\right)
 x_{n}+( 2 b \eta -\beta )  x_{n}^{2}
\label{fnEtaEqNotZeroCase(ii)}
\end{gather}
and the $7$ parameters $a$, $b$, $\alpha$, $\beta$, $\gamma$, $\eta $, $\rho $
are restricted according to 
Table~2.6  (being the
relevant subcase of  Table~2.2):

\begin{figure}[t!]{\centering
{\bf Table 2.6}\label{table2.6}

\vspace{1mm}

$\left\vert
\begin{array}{cccccccc}
\# & a & b & \alpha & \beta & \gamma & \eta & \rho \\
1 & 0 & 0 & \ast & \ast & \ast & \ast & \ast \\
2 & \ast & 0 & \ast & \ast & \ast & \ast & 0 \\
3 & 0 & \ast & b \gamma +\rho & 2 b \eta & \ast & \ast & \ast \\
4 & 0 & \ast & \rho & 2 b \eta & 0 & \ast & \ast \\
5 & \ast & \ast & 2 a \eta & 2 b \eta & \ast & \ast & -b \gamma%
\end{array}%
\right\vert $

}


\vspace{1mm}

{\small \noindent
This table 
indicates the $5$
sets of values to be assigned to the $7$ parameters $a$, $b$, $\alpha$, $\beta$, $\gamma$, $\eta$, $\rho $ in~(\ref{EqMotznEtaNotZero}) with~(\ref{fnEtaEqNotZeroCase(ii)}). Asterisks indicate that the corresponding
parameters can be assigned freely (with $\eta \neq 0$). Note that each of
the f\/irst $2$ lines assigns values to only~$2$ of the $7$ parameters, while
the third and f\/ifth lines assign values to~$3$ of the~$7$ parameters, and
the fourth line assign values to $4$ of the $7$ parameters.}

\end{figure}

These $24$ \textit{Newtonian} equations of motion are exhibited in Appendix~\ref{appendixA}. Their \textit{solvable} character is of course implied by the fact that
the $N$ coordinates $z_{n}(t) $ coincide with the $N$
eigenvalues of the matrix $U(t) $ (see~(\ref{UZ})). To
ascertain the behavior of these solutions $z_{n}(t) $ one must
in each case take account of the restrictions on the parameters
characterizing these models, as detailed above, which are of course also
relevant in order to identify the corresponding evolution of the matrix $%
U(t) $: as implied by inserting in the explicit formula~(\ref{ExplSolUt}) with (\ref{ff})~-- in addition to the parameters $\alpha $, $\beta $, $\gamma$, $\eta $, $\rho $ associated with the \textit{solvable} $%
N $-body model under consideration~-- the expressions of the \textit{initial}
values $U(0) $ and $V(0) \equiv V_{0}$ of the
matrices $U(t) $ and $V(t) $ in terms of the $N$
\textit{initial} values $z_{n}(0) $ of the $N$ coordinates and
the $N$ \textit{initial} values $\dot{z}_{n}(0) $ of the $N$
velocities. To obtain these expressions it is useful to note that it is
possible~-- and convenient~-- to assume that the \textit{diagonalizing} matrix
$R(t) $ (see (\ref{UVM})) is \textit{initially} just the $N\times N$ \textit{unit} matrix~$I$,
\begin{gather}
R(0) =I ,
\end{gather}%
implying \textit{initially} (see (\ref{UVM}))
\begin{subequations}
\begin{gather}
U(0) =\text{diag} [ z_{n}(0)  ]
 , \qquad  U_{nm}(0) =\delta _{nm} z_{n}(0)  ,
\label{Uzero}
\\
V_{0}\equiv V(0) =Y(0)    , \qquad  V_{nm}(0)
=\delta _{nm} y_{n}(0) + ( 1-\delta _{nm} )
 Y_{nm}(0)  .  \label{Vzero}
\end{gather}

The f\/irst of these two formulas provides the explicit expression of $U (0) $ in terms of the initial data $z_{n}(0) $.

In the second formula the \textit{initial} values $y_{n}(0) $ of
the \textit{diagonal} elements of the matrix $V_{0}$ in terms of the initial
coordinates $z_{n}(0) $ and velocities $\dot{z}_{n}(0) $ of the $N$ particles read
\begin{gather}
y_{n}(0) =\frac{\dot{z}_{n}(0) -\alpha  z_{n} (0) -\beta  z_{n}^{2}(0) }{\gamma +2 \eta  z_{n}(0) }  \label{ynzero}
\end{gather}%
(see~(\ref{yn})), while the \textit{off-diagonal} elements $Y_{nm}(0) $ (with $n\neq m$) of the matrix $V_{0}$ are given by the \textit{ansatz}~(\ref{Ans2}) yielding
\begin{gather}
Y_{nm}(0) =\big\{ g_{n}(0)  g_{m}(0)
[ \dot{z}_{n}(0) +f_{n}(0) ]  [ \dot{%
z}_{m}(0) +f_{m}(0) ] \big\}
^{1/2} , \qquad  n\neq m ,  \label{Ynmzerob}
\end{gather}%
with the quantities $g_{n}(0) $ and $f_{n}(0) $
given by the formulas (see (\ref{fg}))%
\begin{gather}
g_{n}(0) =\frac{1}{\gamma +2 \eta  z_{n}(0) }
 ,  \qquad f_{n}=f^{(0) }+f^{(1) } z_{n}(0)
+f^{(2) } z_{n}^{2}(0)  ,
\end{gather}
\end{subequations}
with the appropriate assignments of the parameters $\gamma $, $\eta $, $f^{(0) }$, $f^{(1) }$ and $f^{(2) }$
characterizing the various \textit{solvable} many-body models, see above (in
particular for $f^{(0) }$, $f^{(1) }$ and $f^{(2) }$ see~(\ref{fnEtaEqZeroCase(i)}) or~(\ref{fnEtaEqZeroCase(ii)}) or~(\ref{fnEtaEqNotZeroCase(i)}) or~(\ref{fnEtaEqNotZeroCase(ii)}), as
appropriate).

Note moreover that the \textit{dyadic} character of the \textit{off-diagonal}
part of the matrix $V_{0}=Y(0)$, see (\ref{Ynmzerob}), implies
a simplif\/ication when one must compute functions of this matrix $V_{0}$ such
as those appearing in the explicit expression (\ref{ExplSolUt}) with (\ref{ff}) of $U(t) $; this simplif\/ication becomes particularly
signif\/icant when the matrix $V_{0}=Y(0) $ is altogether \textit{%
dyadic}, $Y_{nm}(0) =v_{n} v_{m}$, since for any \textit{dyadic}
matrix, say $X_{nm}=x_{n} x_{m}$, there holds the simple formula
\begin{gather}
\varphi  ( X ) =\varphi (0)  I+\frac{\varphi  (
x ) -\varphi (0) }{x} X ,  \qquad x^{2}=\sum_{k=1}^{N}x_{n}^{2} ,
\end{gather}%
where $\varphi  ( x ) $ is any (scalar) function for which the
(matrix) expression $\varphi  ( X ) $ makes good sense. This
simplif\/ication clearly happens if\/f
\begin{subequations}
\begin{gather}
f_{n}=-\alpha  z_{n}-\beta  z_{n}^{2} ,
\end{gather}%
as implied by (\ref{yn}) and (\ref{Ans2}) with (\ref{fg}), hence in case~$(i)$ whenever (see (\ref{case(i)}) and  Table~2.1)%
\begin{gather}
a \gamma =0 ,  \qquad 2 a \eta +b \gamma =-\alpha  ,  \qquad 2 b \eta =-\beta  ,
\end{gather}%
and in  case~$(ii)$ whenever (see~(\ref{case(ii)}) and  Table~2.2)
\begin{gather}
a=b=0 .
\end{gather}
\end{subequations}

{\sloppy \textit{Special cases and their $($autonomous$)$ isochronous variants}. Certain
special models among those identif\/ied above as \textit{solvable} can be
\textit{isochronized} via the following change of dependent and independent
variables,
\begin{gather}
z_{n}(t) =\exp  ( \text{\textbf{i}} \sigma  \omega  t )
 \zeta _{n}\left( \tau \right)  ,  \qquad \tau =\frac{\exp  ( \text{\textbf{i}%
} \omega  t ) -1}{\text{\textbf{i}} \omega } .  \label{trick}
\end{gather}
Here the quantities $\zeta _{n} ( \tau  ) $ are assumed to satisfy
the \textit{Newtonian} equations written above, see~(\ref{EqMotznEtaEqZero})
with~(\ref{fnEtaEqZeroCase(i)}) or~(\ref{fnEtaEqZeroCase(ii)}), of course
with the new (\textit{complex}) independent variable $\tau $ repla\-cing the
time~$t$; $\omega $~is an \textit{arbitrary} \textit{real} (for
def\/initeness, \textit{positive}) \textit{constant} to which we associate the
period
\begin{gather}
T=\frac{2 \pi }{\omega } ;  \label{T}
\end{gather}%
and the number $\sigma $ is adjusted so as to produce, for the dependent
variables $z_{n}\equiv z_{n}(t) $ (with the \textit{real}
independent variable $t$ interpreted as ``time'') \textit{autonomous}
equations of motion (the special models providing the starting points for
the application of this trick being appropriately selected to allow such an
outcome). Since the application of this trick is by now quite standard (see,
for instance, Section~2.1 entitled ``The trick'' of~\cite{C2008}), we dispense
here from any detailed discussion of this approach and limit ourselves to
reporting the results.

}

This trick is only applicable to~(\ref{EqMotznEtaEqZero}) with~(\ref{fnEtaEqZeroCase(i)}) in the very special cases with $\alpha =\rho =0$ and
either $\gamma =0$ or $a=b=0$ (as long as one is only interested in getting
\textit{autonomous} equations of motion). Then the assignment $\sigma =1$
yields the \textit{isochronous} equations of motion
\begin{subequations}
\label{Isozn}
\begin{gather}
\ddot{z}_{n} =3 \text{\textbf{i}} \omega  \dot{z}_{n}+2 \omega
^{2} z_{n}+2 \beta  z_{n}  ( \dot{z}_{n}-\text{\textbf{i}} \omega
 z_{n} )  +2  ( \dot{z}_{n}-\text{\textbf{i}} \omega  z_{n} )  \sum_{\ell
=1,\, \ell \neq n}^{N}\left[ \frac{ ( \dot{z}_{\ell }-\text{\textbf{i}}%
 \omega  z_{\ell } ) }{z_{n}-z_{\ell }}\right]  .
\end{gather}

Likewise the application of this trick to (\ref{EqMotznEtaEqZero}) with (\ref{fnEtaEqZeroCase(ii)}), again in the very special cases with $\alpha =\rho
=0 $ and either $\gamma =0$ or $a=b=0$, and again with the assignment $\sigma =1 $, yields the \textit{isochronous} equations of motion%
\begin{gather}
\ddot{z}_{n}  = 3 \text{\textbf{i}} \omega  \dot{z}_{n}+2 \omega
^{2} z_{n}+2 \beta  z_{n}  ( \dot{z}_{n}-\text{\textbf{i}} \omega
 z_{n} )  \nonumber \\
\hphantom{\ddot{z}_{n}  =}{}
+2  ( \dot{z}_{n}-\text{\textbf{i}} \omega  z_{n}-\beta
 z_{n}^{2} )  \sum_{\ell =1, \ell \neq n}^{N}\left[ \frac{( \dot{z}
_{\ell }-\text{\textbf{i}} \omega  z_{\ell }-\beta  z_{\ell }^{2}) }{z_{n}-z_{\ell }}\right]  .
\end{gather}

Neither one of these two models is however new: see Examples 4.2.2-6 and
4.2.2-7 in~\cite{C2008}.

An analogous treatment applied, \textit{mutatis mutandis}, to (\ref%
{EqMotznEtaNotZero}) with (\ref{fnEtaEqNotZeroCase(i)}) in the special case
with $\rho =0$, $2 a \eta =b \gamma $ and $2 \alpha  \eta =\beta  \gamma $
(implying $\lambda =0$), yields (again via the assignment $\sigma =1$) the
\textit{isochronous} equations of motion
\begin{gather}
 \ddot{x}_{n}=\text{\textbf{i}} \omega  \dot{x}_{n}+\omega ^{2} x_{n}+\frac{\dot{x}_{n}^{2}}{x_{n}}+\beta  x_{n} ( \dot{x}_{n}-\text{\textbf{i}}%
 \omega  x_{n})   +\left( \dot{x}_{n}-\text{\textbf{i}} \omega  x_{n}+2 b \eta
 x_{n}^{2}\right)    \nonumber \\
\hphantom{\ddot{x}_{n}=}{}
 \times  \sum_{\ell =1, \ell \neq n}^{N}\left[ \frac{( \dot{x}_{\ell }-
\text{\textbf{i}} \omega  x_{\ell }+2 b \eta  x_{\ell }^{2})  (
x_{n}+x_{\ell }) }{( x_{n}-x_{\ell })  x_{\ell }}\right]  .
\end{gather}

Likewise an analogous treatment applied to (\ref{EqMotznEtaNotZero}) with (\ref{fnEtaEqNotZeroCase(ii)}) in the special case with $\rho =0$ and either $2 \alpha  \eta =\beta  \gamma $ (implying $\lambda =0$) and $\alpha
=-2 a \eta +b \gamma $ or $\gamma =0$ (implying $\lambda =0$) and $\alpha
=2 a \eta $, yields (again via the assignment $\sigma =1$) the same \textit{isochronous} equations of motion (up to the, merely notational, replacement
of $2 b \eta $ with $2 b \eta -\beta $).

While f\/inally this treatment applied, with $\sigma =-1$, to (\ref{EqMotznEtaNotZero}) with (\ref{fnEtaEqNotZeroCase(ii)}) in the special case
with $b=\beta =\rho =0$ and $\alpha =2 a \eta $ yields the \textit{isochronous} equations of motion
\begin{gather}
\ddot{x}_{n}=\text{\textbf{i}} \omega  \dot{x}_{n}-\omega ^{2} x_{n}+\frac{%
\dot{x}_{n}^{2}}{x_{n}}+a \gamma  \frac{\dot{x}_{n}}{x_{n}}+a \gamma  \text{%
\textbf{i}} \omega    \nonumber \\
\hphantom{\ddot{x}_{n}=}{}
+ ( \dot{x}_{n}+\text{\textbf{i}} \omega  x_{n}+a \gamma  )
 \sum_{\ell =1, \ell \neq n}^{N}\left[ \frac{ ( \dot{x}_{\ell }+\text{\textbf{i}} \omega  x_{\ell }+a \gamma  )  ( x_{n}+x_{\ell
} ) }{( x_{n}-x_{\ell })  x_{\ell }}\right]  .
\end{gather}
\end{subequations}

\subsection{A related class of \textit{solvable} many-body models}\label{section2.3}

In this subsection we consider the \textit{Newtonian} equations of motion
that obtain by identifying the~$N$ dependent variables of the models
discussed above as the $N$ \textit{zeros} of a monic (time-dependent)
polynomial of degree~$N$, and by then focussing on the time-evolution of the
$N$ \textit{coefficients} of this polynomial. It is again convenient to
treat separately the two cases with $\eta =0$ and with $\eta \neq 0$.

In the $\eta =0$ case the starting point are the equations of motion (\ref{EqMotznEtaEqZero}) with (\ref{fnEtaEqZeroCase(i)}) or (\ref{fnEtaEqZeroCase(ii)}). We then introduce the time-dependent (monic)
polynomial $\psi ( z,t) $ whose zeros are the $N$ eigenvalues $z_{n}(t) $ of the $N\times N$ matrix $U(t) $:
\begin{subequations}\label{Psi}
\begin{gather}
\psi  ( z,t ) =\det  [ z I-U(t)  ]  ,
\\
\psi \left( z,t\right) =\prod_{n=1}^{N} [ z-z_{n}(t)  ]
=z^{N}+\sum_{m=1}^{N} \big[ c_{m}(t)  z^{N-m} \big]  .
\label{psipsi}
\end{gather}
The last of these formulas introduces the $N$ coef\/f\/icients $c_{m}\equiv
c_{m}(t) $ of the monic polynomial $\psi  ( z,t ) $; of
course it implies that these coef\/f\/icients are related to the zeros $z_{n}(t) $ as follows:
\begin{gather}
c_{1}=-\sum_{n=1}^{N}z_{n} ,   \qquad c_{2}=\sum_{n,m=1; n>m}^{N}z_{n} z_{m} ,
\label{c12}
\end{gather}
\end{subequations}
and so on.

The fact that the initial-value problem associated with the time evolution
of the $N$ coordinates $z_{n}$ can be \textit{solved} by \textit{algebraic}
operations implies that the same \textit{solvable} character can be
attributed to the time evolution of the monic polynomial $\psi  (
z,t ) $ and of the $N$ coef\/f\/icients~$c_{m}(t) $. The
procedure to obtain the equations of motion satisf\/ied by the~$N$
coef\/f\/icients~$c_{m}(t) $ from the~$N$ equations of motion
satisf\/ied by the $N$ zeros is tedious but standard; a key role in this
deve\-lop\-ment are the identities reported, for instance, in Appendix~A of \cite{C2008} (but note that there are two misprints in these formulas: in equation~(A.8k) the term $ ( N+1 ) $ inside the square brackets should
instead read $ ( N-3 ) $; in equation~(A.8l) the term $N^{2}$ inside the
square brackets should instead read $N ( N-2 ) $~-- these
misprints have been corrected in the recent paperback version of this
monograph~\cite{C2008}). Here we limit
our presentation to reporting the f\/inal result.

The equation characterizing the time evolution of the monic polynomial $\psi
 ( z,t ) $ implied by the \textit{Newtonian} equations of motion (\ref{EqMotznEtaEqZero}) with (\ref{fn}) reads as follows:
\begin{gather}
\psi _{tt}-2 \big( f^{(0) }+f^{(1) } z+f^{(2) } z^{2}\big)  \psi _{zt}+\big( p^{(0) }+p^{(1) } z\big)  \psi _{t}   \nonumber\\
\qquad{}
+\big( q_{2}^{(0) }+q_{2}^{(1) } z+q_{2}^{(2) } z^{2}+q_{2}^{( 3) } z^{3}+q_{2}^{( 4)
} z^{4}\big)  \psi _{zz}    \nonumber \\
\qquad{} +\big( q_{1}^{(0) }+q_{1}^{(1) } z+q_{1}^{(2) } z^{2}+q_{1}^{( 3) } z^{3}\big)  \psi _{z}
 +\big( q_{0}^{(0) }+q_{0}^{(1) } z+q_{0}^{(2) } z^{2}\big)  \psi =0 ,    \label{PDEpsi}
\end{gather}
with
\begin{subequations}
\label{pq}
\begin{gather}
p^{(0) }=-\alpha -\rho +2  ( N-1 )  f^{(1)
}-2 f^{(2) } c_{1} ,  \qquad  p^{(1) }=2 \big[ {-}\beta
+ ( N-2 )  f^{(2) }\big]  ;
\\
q_{2}^{(0) }  = \big( f^{(0) }\big)
^{2} , \qquad  q_{2}^{(1) }=2 f^{(0) } f^{(1)
} ,  \qquad  q_{2}^{(2) }=2 f^{(0) } f^{(2)
}+\big( f^{(1) }\big) ^{2} ,  \nonumber \\
q_{2}^{(3) }  = 2 f^{(1) } f^{(2)
} ,  \qquad  q_{2}^{(4) }=\big( f^{(2) }\big) ^{2} ;
\\
q_{1}^{(0) }  = -2 (N-1)  f^{(0)
} f^{(1) }+2 f^{(0) } f^{(2) } c_{1} ,
\nonumber \\
q_{1}^{(1) }  = -\alpha  \rho -2 (N-2)  f^{(0) } f^{(2) }+2 f^{(1) } f^{(2)
} c_{1} -2 (N-1)  \big( f^{(1) }\big) ^{2} ,
\nonumber \\
q_{1}^{(2) }  = -\beta  \rho -2  ( 2 N-3 )  f^{ (1) } f^{(2) }+2 \big( f^{(2) }\big)
^{2} c_{1} ,  \nonumber \\
q_{1}^{(3) }  = -2 (N-2)  \big( f^{(2)
}\big) ^{2} ;
\\
q_{0}^{(0) }  = N \alpha  \rho -2 N f^{(0)
} f^{(2) }+N (N-1)  \big( f^{(1)
}\big) ^{2}  \nonumber \\
\hphantom{q_{0}^{(0) }  =}{}
 -\big[ \beta  \rho +2 (N-1)  f^{(1) } f^{(2) }\big]  c_{1}+2 \big( \beta +f^{(2) }\big)  \dot{c}%
_{1}+2 \big( f^{(2) }\big) ^{2} c_{2} ,  \nonumber \\
q_{0}^{(1) }  = N \beta  \rho +2 N (N-2)  f^{(1) } f^{(2) }-2 (N-1)  \big( f^{(2) }\big) ^{2} c_{1} ,  \nonumber \\
q_{0}^{(2) }  = N  ( N-3 )  \big( f^{(2)
}\big) ^{2} ,
\end{gather}
\end{subequations}
where of course the quantities $f^{(0) }$, $f^{(1)
}$, $f^{(2) }$ should be expressed in terms of the other
parameters as implied by~(\ref{fn}) with~(\ref{fnEtaEqZeroCase(i)}) or~(\ref{fnEtaEqZeroCase(ii)}), as the case may be. Note that, while this equa\-tion,~(\ref{PDEpsi}), satisf\/ied by the function $\psi  ( z,t ) $ (where of
course subscripted variables denote partial dif\/ferentiations) might seem a
\textit{linear PDE}, it is in fact a \textit{nonlinear functional equation},
because some of its coef\/f\/icients, see~(\ref{pq}), depend on the quantities $c_{1}$ and $c_{2}$ which themselves depend on~$\psi $, indeed clearly (see~(\ref{Psi}))
\begin{subequations}
\begin{gather}
c_{m}\equiv c_{m}(t) =\frac{\psi ^{ ( N-m ) }(0,t) }{( N-m) !} ,
\end{gather}%
where we used the shorthand notation $\psi ^{( j) }(z,t) $ to denote the $j$-th partial derivative with respect to the
variable $z$ of $\psi ( z,t) $,
\begin{gather}
\psi ^{( j) }( z,t) \equiv \frac{\partial ^{j} \psi
( z,t) }{\partial  z^{j}} , \qquad  j=1,2,\dots .
\end{gather}
\end{subequations}

Likewise, the equation characterizing the time evolution of the monic
polynomial $\phi \left( x,t\right) $ implied by the \textit{Newtonian}
equations of motion (\ref{EqMotznEtaNotZero}) via the following assignment
(analogous to (\ref{psipsi})),
\begin{gather}
\phi  ( x,t ) =\prod_{n=1}^{N} [ x-x_{n}(t)  ]
=x^{N}+\sum_{m=1}^{N}\big[ c_{m}(t)  x^{N-m}\big]  ,
\label{chi}
\end{gather}
reads
\begin{gather}
\phi _{tt}-2 \big( f^{(0) }+f^{(1) } x+f^{(2) } x^{2}\big)  \phi _{xt}+\left( \frac{p^{( -1) }}{x}
+p^{(0) }+p^{(1) } x\right)  \phi _{t}    \nonumber
\\
\qquad+\big( q_{2}^{(0) }+q_{2}^{(1) } x+q_{2}^{(2) } x^{2}+q_{2}^{(3) } x^{3}+q_{2}^{(4)
} x^{4}\big)  \phi _{xx}     \\
\qquad{}
+\left( \frac{q_{1}^{( -1) }}{x}+q_{1}^{(0)
}+q_{1}^{(1) } x+q_{1}^{(2) } x^{2}+q_{1}^{(3) } x^{3}\right)  \phi _{x}   \nonumber \\
\qquad{}+\left( \frac{q_{0}^{( -1) }}{x}+q_{0}^{(0)
}+q_{0}^{(1) } x+q_{0}^{(2) } x^{2}\right)  \phi
=0,    \label{PDEchi}
\end{gather}
now with
\begin{subequations}
\label{pqchi}
\begin{gather}
p^{\left( -1\right) }  = 2 f^{(0) }-\lambda  ,  \qquad
p^{(0) }  = -\rho +N f^{(1) }-f^{(2)
} c_{1}+f^{(0) } \frac{c_{N-1}}{c_{N}}-\frac{\dot{c}_{N}}{c_{N}}%
 ,  \nonumber \\
p^{(1) }  = 2 (N-1)  f^{(2) }-\beta  ;
\\
q_{2}^{(0) }  = \big( f^{(0) }\big)
^{2} ,   \qquad q_{2}^{(1) }=2 f^{(0) } f^{(1)
} ,   \qquad q_{2}^{(2) }=2 f^{(0) } f^{(2)
}+\big( f^{(1) }\big) ^{2} ,  \nonumber \\
q_{2}^{(3) }  = 2 f^{(1) } f^{(2)
} ,   \qquad q_{2}^{(4) }=\big( f^{(2) }\big) ^{2} ;
\\
q_{1}^{( -1) }  = -\big(  f^{(0) }\big) ^{2}  ,
\qquad
q_{1}^{(0) }  = \rho  \lambda -N f^{(0) } f^{(1) }+f^{(0) } f^{(2) } c_{1}+f^{(0) } \frac{\dot{c}_{N}}{c_{N}}-\big( f^{(0) }\big) ^{2} %
\frac{c_{N-1}}{c_{N}} ,  \nonumber \\
q_{1}^{(1) }  = -\rho  \mu -2 (N-1)  f^{(0) } f^{(2) }+f^{(1) } f^{(2)
} c_{1}-(N-1)  \big( f^{(1) }\big) ^{2}
 +f^{(1) } \frac{\dot{c}_{N}}{c_{N}}-f^{(0)
} f^{(1) } \frac{c_{N-1}}{c_{N}} ,  \nonumber \\
q_{1}^{(2) }  = -\beta  \rho - ( 3 N-4 )  f^{(1 ) } f^{(2) }
 +\big( f^{(2) }\big) ^{2} c_{1}+f^{(2) } \frac{\dot{c}_{N}}{c_{N}}-f^{(0) } f^{(2) } \frac{c_{N-1}}{%
c_{N}} ,  \nonumber \\
q_{1}^{(3) }  = - ( 2 N-3 )  \big( f^{(2)
}\big) ^{2} ;
\\
q_{0}^{( -1) }  = \big( \lambda -2 f^{(0) }\big)  %
\frac{\dot{c}_{N}}{c_{N}}+\big( f^{(0) }\big) ^{2} \frac{%
c_{N-1}}{c_{N}} ,  \nonumber \\
q_{0}^{(0) }  = -\beta  \rho  c_{1}+\beta  \dot{c}_{1}+\big(
f^{(2) } c_{1}-N f^{(1) }\big)  \frac{\dot{c}_{N}}{%
c_{N}}
 +f^{(0) } \big( N f^{(1) }-f^{(2)
} c_{1}\big)  \frac{c_{N-1}}{c_{N}} ,  \nonumber \\
q_{0}^{(1) }  = N \beta  \rho +N (N-2)  f^{(1) } f^{(2) }-(N-1)  \big( f^{(2)
}\big) ^{2} c_{1}    -N f^{(2) } \frac{\dot{c}_{N}}{c_{N}}+f^{(0)
} f^{(2) } \frac{c_{N-1}}{c_{N}} ,  \nonumber \\
q_{0}^{(2) }  = N (N-2)  \big( f^{(2)
}\big) ^{2} .
\end{gather}
\end{subequations}

The equations of motion of \textit{Newtonian} type satisf\/ied by the $N$
coef\/f\/icients $c_{m}(t) $ which obtain from (\ref{PDEpsi}) hence
correspond to the \textit{Newtonian} equations of motion (\ref{EqMotznEtaEqZero}) read as follows:
\begin{gather}
\ddot{c}_{m}-2  ( N-m+1 )  f^{(0) } \dot{c}_{m-1}+\big[
-2  ( N-m )  f^{(1) }+p^{(0) }\big]  \dot{c}_{m} \nonumber\\
\qquad+\big[ {-}2  ( N-m-1 )  f^{(2) }+p^{(1) }\big]  \dot{c}_{m+1}
+ ( N-m+2 )   ( N-m+1 )  q_{2}^{(0) } c_{m-2}
  \nonumber \\
\qquad{}+ ( N-m+1 )  \big[  ( N-m )  q_{2}^{(1)
}+q_{1}^{(0) }\big]  c_{m-1}  \nonumber \\
\qquad{}
+\big\{  ( N-m )  \big[  ( N-m-1 )  q_{2}^{(2) }+q_{1}^{(1) }\big] +q_{0}^{(0) }\big\}
 c_{m}    \nonumber \\
\qquad{}
+\big\{  ( N-m-1 )  \big[  ( N-m-2 )  q_{2}^{(3) }+q_{1}^{(2) }\big] +q_{0}^{(1) }\big\}
 c_{m+1}    \nonumber \\
\qquad {} +\big\{  ( N-m-2 )  \big[  ( N-m-3 )  q_{2}^{(4) }+q_{1}^{(3) }\big] +q_{0}^{(2) }\big\}
 c_{m+2}=0 ,    \label{Eqcm}
\end{gather}
where $c_{n}$ vanishes for $n<0$ and for $n>N$ while $c_{0}=1$ (see~(\ref{Psi})), and of course the coef\/f\/icients~$p^{( j) }$ and~$q_{k}^{( j) }$ are def\/ined by~(\ref{pq}). Again, this system of
ODEs might seem \textit{linear}, but it is in fact \textit{nonlinear}
because some of its coef\/f\/icients depend on the dependent variables~$c_{1}$
and~$c_{2}$, see~(\ref{pq}). The more explicit version of these equations of
motion that obtain by expressing the various coef\/f\/icients in terms of the
free parameters are listed in Appendix~\ref{appendixB}. They are of course just as \textit{solvable} as the \textit{Newtonian} equations of motion satisf\/ied by the~$N$
coordinates~$x_{n}(t) $, see Appendix~\ref{appendixA}, to which they
correspond via~(\ref{Psi}); and in particular whenever the parameter~$\rho $
is \textit{imaginary} and the parameter~$\alpha $ is \textit{real} and
\textit{negative} they are \textit{asymptotically isochronous} with period~$T$ (see Remark~\ref{remark2.3}).

Likewise, the equations of motion of \textit{Newtonian} type satisf\/ied by
the $N$ coef\/f\/icients $c_{m}(t) $ which obtain from (\ref{PDEchi}) hence correspond to the \textit{Newtonian} equations of motion (\ref{EqMotznEtaNotZero}) read as follows:
\begin{gather}
\ddot{c}_{m}+\big[ p^{( -1) }-2 ( N-m+1)  f^{(0) }\big]  \dot{c}_{m-1}
+\big[ {-}2  ( N-m )  f^{(1) }+p^{(0) }
\big]  \dot{c}_{m}\nonumber\\
\qquad{}+\big[ {-}2  ( N-m-1 )  f^{(2)
}+p^{(1) }\big]  \dot{c}_{m+1}
+ ( N-m+2 )  \big[  ( N-m+1 )  q_{2}^{(0)
}+q_{1}^{ ( -1 ) }\big]  c_{m-2}    \nonumber \\
\qquad {}
+\big\{  ( N-m+1 )  \big[  ( N-m )  q_{2}^{(1) }+q_{1}^{(0) }\big] +q_{0}^{( -1)
}\big\}  c_{m-1}   \nonumber \\
\qquad {}
+\big\{  ( N-m )  \big[  ( N-m-1 )  q_{2}^{(2) }+q_{1}^{(1) }\big] +q_{0}^{(0) }\big\}
 c_{m}    \nonumber \\
\qquad{}
+\big\{  ( N-m-1 )  \big[  ( N-m-2 )  q_{2}^{(3) }+q_{1}^{(2) }\big] +q_{0}^{(1) }\big\}
 c_{m+1}    \nonumber \\
 \qquad{}
+\big\{  ( N-m-2 )  \big[  ( N-m-3 )  q_{2}^{ (4) }+q_{1}^{(3) }\big] +q_{0}^{(2) }\big\}
 c_{m+2}=0 ,    \label{Eqcm2}
\end{gather}
where of course again $c_{n}$ vanishes for $n<0$ and for $n>N$ while $c_{0}=1 $ (see (\ref{chi})) and of course the coef\/f\/icients $p^{(j)}$ and $q_{k}^{(j) }$ are now def\/ined by~(\ref{pqchi}).
Again, this system of ODEs might seem \textit{linear}, but it is in fact
\textit{nonlinear} because some of its coef\/f\/icients depend on the dependent
variables $c_{1}$, $c_{N-1}$ and $c_{N}$, see~(\ref{pqchi}). The more
explicit version of these equations of motion that obtains by expressing the
various coef\/f\/icients in terms of the free parameters are listed in Appendix~\ref{appendixB}. They are of course just as \textit{solvable} as the \textit{Newtonian}
equations of motion satisf\/ied by the $N$ coordinates $x_{n}(t) $, see Appendix~\ref{appendixA}, to which they correspond via~(\ref{chi}); and in
particular whenever the parameter $\rho $ is \textit{imaginary }and the
parameter $\alpha $ is \textit{real} and \textit{negative} they are \textit{asymptotically isochronous} with period $T$ (see  Remark~\ref{remark2.3}).

\textit{Special cases and their (autonomous) isochronous variants}. Certain
special models among those identif\/ied above (in this subsection) as
\textit{solvable} can be \textit{isochronized} by an analogous trick to that
employed at the end of the preceding Subsection~\ref{section2.2}. One route to this end
takes as starting point the \textit{isochronized} systems of \textit{Newtonian} equations of motion (\ref{Isozn}) and applies to them the same
procedure employed above to obtain the equations of motion (\ref{Eqcm}) with
(\ref{pq}) and (\ref{Eqcm2}) with (\ref{pqchi}). An equivalent procedure is
to apply to certain special subcases of these systems of ODEs, (\ref{Eqcm})
with (\ref{pq}) and (\ref{Eqcm2}) with (\ref{pqchi}), the following change
of dependent and independent variables:%
\begin{gather}
c_{m}(t) =\exp \left( \text{\textbf{i} }\sigma  m \omega
 t\right)  \chi _{m}\left( \tau \right)  ,   \tau =\frac{\exp \left( \text{%
\textbf{i}} \omega  t\right) -1}{\text{\textbf{i}} \omega } .
\end{gather}%
Here the quantities $\chi _{n}\left( \tau \right) $ are assumed to satisfy
the systems of ODEs written above, see (\ref{Eqcm}) with (\ref{pq}) and (\ref%
{Eqcm2}) with (\ref{pqchi}), of course with the new (\textit{complex})
independent variable $\tau $ replacing the time $t$; $\omega $ is an \textit{%
arbitrary} \textit{real} (for def\/initeness, \textit{positive}) \textit{%
constant} to which we associate the period $T,$ see (\ref{T}); and the
number $\sigma $ is adjusted so as to produce \textit{autonomous} ODEs for
the new dependent variables $c_{m}\equiv c_{m}(t) $ (with the
\textit{real} independent variable $t$ interpreted as ``time'': the special
models providing the starting points for the application of this trick being
appropriately selected in order to allow such an outcome). Since the
application of this trick is quite standard, we dispense here from any
detailed discussion of this approach and limit ourselves to reporting the
results.

The ODEs that follow from (\ref{Eqcm}) (with $\alpha =\rho =\eta =0,$ $%
f^{(0) }=f^{(1) }=f^{(2) }=0$ and $%
\sigma =1$) read as follows:
\begin{subequations}
\label{Isocm}
\begin{gather}
\ddot{c}_{m}  = \text{\textbf{i}} (2 m+1) \omega  \dot{c}_{m}-\big[2 \beta  (
\dot{c}_{1}-\text{\textbf{i}} \omega  c_{1})-m (m+1) \omega ^{2}\big] c_{m}
\nonumber \\
 \hphantom{\ddot{c}_{m}  =}{}
 +2 \beta  \dot{c}_{m+1}-\text{\textbf{i}} 2 \beta  (m+1) \omega  c_{m+1} ;
\end{gather}
those that follow from (\ref{Eqcm}) (with $\alpha =\rho =\eta =0$, $f^{(0) }=f^{(1) }=0$, $f^{(2) }=-\beta $
and $\sigma =1$) read instead as follows
\begin{gather}
\ddot{c}_{m}  = [\text{\textbf{i}} (2 m+1) \omega -2 \beta  c_{1}] \dot{c} _{m}   +
\big[m (m+1) \omega ^{2}+\text{\textbf{i}} 2 m \beta  \omega  c_{1}-2 \beta
^{2} c_{2}\big] c_{m}  \nonumber \\
\hphantom{\ddot{c}_{m}  =}{}
+2 m \dot{c}_{m+1}+2 m \big[\beta ^{2} c_{1}-\text{\textbf{i}} (m+1) \omega
\big] c_{m+1}   -(m-1) (m+2) \beta ^{2} c_{m+2} .
\end{gather}
\end{subequations}
Neither one of these two \textit{isochronous} many-body problems is new.

The analogous results that follows from (\ref{Eqcm2}) are instead generally
\textit{new}. There are then two sets of cases. The f\/irst set of \textit{isochronous} models obtain from the assignment $\sigma =1$ and read
\begin{subequations}
\label{Isocm1}
\begin{gather}
\ddot{c}_{m}=\left[-\text{\textbf{i}} (N-2 m-1) \omega +f^{(2)} c_{1}+\frac{\dot{c%
}_{N}}{c_{N}}\right] \dot{c}_{m}
+\big(\beta -2 m f^{(2)}\big) \dot{c}_{m+1}\nonumber\\
\hphantom{\ddot{c}_{m}=}{}
+\left[-m (N-m-1) \omega ^{2}+(N-m) \omega
 f^{(2)} c_{1}
-\frac{\dot{c}_{N}}{c_{N}} \big(\text{\textbf{i}} m \omega +f^{(2)} c_{1}+\beta
 (\text{\textbf{i}} \omega  c_{1}-\dot{c}_{1})\big)\right] c_{m}    \nonumber \\
\hphantom{\ddot{c}_{m}=}{}
+\left[ -\text{\textbf{i}}(N-2 m) (m+1) \omega  f^{(2)}-\text{\textbf{i}}
 (m+1) \omega  \beta
  +m \big(f^{(2)}\big)^{2} c_{1}+(m+1) f^{(2)} \frac{\dot{c}_{N}}{c_{N}}%
\right]  c_{m+1}    \nonumber \\
\hphantom{\ddot{c}_{m}=}{}
-m (m+2) \big(f^{(2)}\big)^{2} c_{m+2} ,
\end{gather}
with the following restriction on the parameters:
\begin{gather}
\rho =\lambda =f^{(0)}=f^{(1)}=0 .
\end{gather}
\end{subequations}

The second set of \textit{isochronous} models obtain from the assignment $%
\sigma =-1$ and read
\begin{subequations}
\label{Isocm2}
\begin{gather}
\ddot{c}_{m}=(2 N-2 m+1) \lambda  \dot{c}_{m-1}
+\left[\text{\textbf{i}} (N-2 m+1) \omega +\frac{\dot{c}_{N}}{c_{N}}-\lambda  %
\frac{c_{N-1}}{c_{N}}\right] \dot{c}_{m}\nonumber \\
\hphantom{\ddot{c}_{m}=}{}
-(N-m) (N-m+2) \lambda ^{2} c_{m-2}
+\bigg[ \text{\textbf{i}} (m-1) (2 N-2 m+1) \omega \nonumber \\
\hphantom{\ddot{c}_{m}=}{}
+(N-m+1) (N-m-1) \lambda  \left(\lambda  \frac{c_{N-1}}{c_{N}}-\text{\textbf{i}} N \omega -\frac{\dot{c}_{N}}{c_{N}}\right)\bigg]  c_{m-1}
\nonumber \\
\hphantom{\ddot{c}_{m}=}{}
+\left[-m (N-m+1) \omega ^{2}+\text{\textbf{i}} m \omega  \left(\frac{\dot{c}_{N}}{%
c_{N}}-\lambda  \frac{c_{N-1}}{c_{N}}\right)\right] c_{m} ,
\end{gather}
with the following restrictions on the parameters:
\begin{gather}
b=\beta =\rho =f^{(1)}=f^{(2)}=0 , \qquad   \alpha =2 a \eta , \qquad f^{(0)}=\lambda
=a \gamma  .
\end{gather}
\end{subequations}

\section{Outlook}\label{section3}

Results analogous, but somewhat more general, than those reported in this
paper can be obtained by an analogous treatment based on a somewhat more
general~-- but still \textit{solvable}~-- system of two $N\times N$
matrix ODEs than~(\ref{EqUV}), such as, for instance,
\begin{gather}
\dot{U}=\alpha  U+\beta  U^{2}+\gamma  V+\eta   ( U V+V U )  ,   \qquad
\dot{V}=\rho _{0}+\rho  V +\rho _{2} V^{2} ,
\end{gather}
which contains the $2$ additional scalar constants $\rho _{0}$ and $\rho
_{2} $ (and clearly reduces to~(\ref{EqUV}) for $\rho _{0}=\rho _{2}=0$).
These developments will be reported in subsequent papers.

Finally, let us recall that \textit{Diophantine} f\/indings can be obtained
from a \textit{nonlinear} \textit{autonomous} \textit{isochronous} dynamical
system by investigating its behavior in the \textit{infinitesimal vicinity}
of its equilibria. The relevant equations of motion become then generally
\textit{linear}, but they of course retain the properties to be \textit{%
autonomous} and \textit{isochronous}. For a system of \textit{linear}
\textit{autonomous} ODEs, the property of \textit{isochrony} implies that
\textit{all} the eigenvalues of the matrix of its coef\/f\/icients are \textit{%
integer numbers} (up to a common rescaling factor). When the \textit{linear}
system describes the behavior of a \textit{nonlinear} \textit{autonomous}
system in the \textit{infinitesimal vicinity} of its equilibria, these
matrices can generally be \textit{explicitly} computed in terms of the
values at equilibrium of the dependent variables of the original, \textit{%
nonlinear} model. In this manner nontrivial \textit{Diophantine} f\/indings
and conjectures have been discovered and proposed: see for instance the
review of such developments in Appendix C (entitled ``Diophantine f\/indings
and conjectures'') of \cite{C2008}. Analogous results obtained by applying
this approach to the \textit{isochronous} systems of \textit{autonomous
nonlinear} ODEs introduced above~-- and in subsequent papers~-- will be
reported if they turn out to be novel and interesting.

\appendix

\section{First appendix}\label{appendixA}

In this appendix we list the $24$ \textit{Newtonian} equations of
motion whose \textit{solvable} character has been demonstrated in this
paper. In each case the parameters they feature (such as $a$, $b$, $\alpha$,
$\beta$, $\gamma$, $\eta$, $\rho $, as the case may be) are \textit{arbitrary constants}; the (assigned) values of the other ones of these
parameters (which also characterize the time-evolution of the solutions of
these equations, see~(\ref{ExplSolUt})), are also reported. Let us emphasize
that if the parameter $\rho $ is an \textit{imaginary} number and the
parameter $\alpha $ is \textit{real} and \textit{negative}, the
corresponding many-body problem is \textit{asymptotically isochronous} with
period $T$, see  Remark~\ref{remark2.3}; and that \textit{isochronous} many-body
models are characterized by the $4$ \textit{Newtonian} equations of motion~(\ref{Isozn}) displayed at the end of Subsection~\ref{section2.2}. Let us also mention
again that the equations of motion reported below are \textit{not} all new;
in particular \textit{not} new are clearly those whose corresponding
equations of motion in the following Appendix~\ref{appendixB} are \textit{linear}.

$\eta =0$,  case~$(i)$  ($5$ models, corresponding to  Table~2.3):

(1) $a=\beta =0$,   $\alpha =-b \gamma$,   $f_{n}=-\alpha  z_{n}$:
\begin{subequations}
\begin{gather}
\ddot{z}_{n}=-\alpha  \rho  z_{n}+ ( \alpha +\rho  )  \dot{z}
_{n}+2  ( \dot{z}_{n}-\alpha  z_{n} )  \sum_{\ell =1,\,\ell \neq
n}^{N}\left( \frac{\dot{z}_{\ell }-\alpha  z_{\ell }}{z_{n}-z_{\ell }}
\right)  ;
\end{gather}

(2) $\alpha =-b \gamma$,   $\beta =\rho =0$,   $f_{n}=- ( a/b )
 \alpha -\alpha  z_{n}$:
\begin{gather}
\ddot{z}_{n}=-\alpha  \dot{z}_{n}+2 \left( \dot{z}_{n}-\frac{a \alpha }{b}
-\alpha  z_{n}\right)  \sum_{\ell =1,\, \ell \neq n}^{N}\left( \frac{\dot{z}
_{\ell }-a \alpha /b-\alpha  z_{\ell }}{z_{n}-z_{\ell }}\right)  ;
\end{gather}

(3) $a=\beta =0$,   $\rho =-b \gamma$,   $f_{n}=-\rho  z_{n}$:
\begin{gather}
\ddot{z}_{n}=-\alpha  \rho  z_{n}+ ( \alpha +\rho  )  \dot{z}
_{n}+2  ( \dot{z}_{n}-\rho  z_{n} )  \sum_{\ell =1,\, \ell \neq
n}^{N}\left( \frac{\dot{z}_{\ell }-\rho  z_{\ell }}{z_{n}- z_{\ell }}\right)
 ;
\end{gather}

(4) $a=0$,   $\alpha =-b \gamma$,   $\rho =-2 b \gamma =2 \alpha$,   $f_{n}=-\alpha  z_{n}$:
\begin{gather}
\ddot{z}_{n}=-2 \alpha ^{2} z_{n}-2 \alpha  \beta  z_{n}^{2}+3 \alpha  \dot{z}_{n}+ 2 \beta  \dot{z}_{n}z_{n}+2  ( \dot{z}_{n}-\alpha  z_{n} )
 \sum_{\ell =1,\ell \neq n}^{N}\left( \frac{\dot{z}_{\ell }- \alpha  z_{\ell
}}{z_{n}-z_{\ell }}\right)  ;
\end{gather}

(5) $\alpha =b \gamma$,   $\beta =b^{2} \gamma /a=b \alpha /a$,   $\rho
=-2 b \gamma =-2 \alpha$,   $f_{n}=a \alpha /b+\alpha  z_{n}$:
\begin{gather}
 \ddot{z}_{n}=2 \alpha ^{2} z_{n}+\frac{2 b}{a} \alpha ^{2} z_{n}^{2}+\frac{%
2 b}{a} \alpha  \dot{z}_{n} z_{n}-\alpha  \dot{z}_{n}   \nonumber \\
\hphantom{\ddot{z}_{n}=}{}
+2 \left( \dot{z}_{n}+\alpha  \frac{a}{b}+\alpha  z_{n}\right)  \sum_{\ell
=1,\ell \neq n}^{N}\left( \frac{\dot{z}_{\ell }+a \alpha /b+ \alpha  z_{\ell
}}{z_{n}-z_{\ell }}\right)  .
\end{gather}
\end{subequations}

$\eta =0$,  case $(ii)$ ($2$ models, corresponding to Table~2.4):

(1) $a=\beta =0$,   $\alpha =b \gamma +\rho$,   $f_{n}=-\rho  z_{n}$:
\begin{subequations}
\begin{gather}
\ddot{z}_{n}=- ( b \gamma +\rho  )  \rho  z_{n}+ ( b \gamma
+2 \rho  )  \dot{z}_{n}+2  ( \dot{z}_{n}-\rho  z_{n} )
 \sum_{\ell =1,\, \ell \neq n}^{N}\left( \frac{\dot{z}_{\ell }-\rho  z_{\ell }}{z_{n}-z_{\ell }}\right)  ;
\end{gather}

(2) $\alpha =\beta =0$,   $\rho =-b \gamma$,   $f_{n}=a \gamma +b \gamma  z_{n}$:
\begin{gather}
\ddot{z}_{n}=-b \gamma  \dot{z}_{n}+2  ( \dot{z}_{n}+ a \gamma
+ b \gamma  z_{n} )  \sum_{\ell =1,\, \ell \neq n}^{N}\left( \frac{\dot{z} _{\ell }+a \gamma + b \gamma z_{\ell }}{z_{n}-z_{\ell }}\right)  .
\end{gather}
\end{subequations}

$\eta \neq 0$,  case $(i)$  ($12$ models, corresponding to  Table~2.5):

(1) $a=b=\rho =0$,   $f_{n}=0$, $ \lambda =(2 \alpha  \eta - \beta  \gamma
) \gamma / ( 4 \eta ^{2} )$, $ \mu =\alpha -\beta  \gamma /\eta$:
\begin{subequations}
\begin{gather}
\ddot{x}_{n}=\frac{\dot{x}_{n}^{2}}{x_{n}}+\lambda \frac{\dot{x}_{n}}{x_{n}}
+\beta  \dot{x}_{n} x_{n}+\dot{x}_{n} \sum_{\ell =1,\, \ell \neq n}^{N}\left[
\frac{\dot{x}_{\ell } ( x_{n}+x_{\ell }) }{( x_{n}-x_{\ell
})  x_{\ell }}\right] ;
\end{gather}

(2) $b=\beta =\gamma =\rho =0$,   $f_{n}=2 a \eta  x_{n}$, $ \lambda =0$, $\mu =\alpha $:
\begin{gather}
\ddot{x}_{n}=\frac{\dot{x}_{n}^{2}}{x_{n}}+( \dot{x}_{n}+2 a \eta
 x_{n})  \sum_{\ell =1,\, \ell \neq n}^{N}\left[ \frac{ ( \dot{x}
_{\ell }+ 2 a \eta  x_{\ell } )   ( x_{n}+x_{\ell } ) }{ (
x_{n}-x_{\ell } )  x_{\ell }}\right]  ;
\end{gather}

(3) $b=\beta =\gamma =0$,   $\alpha =-2 a \eta$,   $f_{n}=-\alpha  x_{n}$, $\lambda =0$, $\mu =\alpha $:
\begin{gather}
\ddot{x}_{n}=\frac{\dot{x}_{n}^{2}}{x_{n}}+\rho   ( \dot{x}_{n}-\alpha
 x_{n} ) + ( \dot{x}_{n}-\alpha  x_{n} )  \sum_{\ell =1,\, \ell
\neq n}^{N}\left[ \frac{(\dot{x}_{\ell }-\alpha  x_{\ell })(x_{n}+x_{\ell })%
}{(x_{n}-x_{\ell }) x_{\ell }}\right]  ;
\end{gather}

(4) $b=\beta =0$,   $\alpha =-2 a \eta$,   $\rho =-\alpha$,   $f_{n}=-\alpha
 x_{n}$, $ \lambda =\alpha  \gamma / ( 2 \eta  )$, $  \mu
=\alpha $:
\begin{gather}
 \ddot{x}_{n}=\frac{\dot{x}_{n}^{2}}{x_{n}}+\frac{\alpha  \gamma }{2 \eta }
\frac{\dot{x}_{n}}{x_{n}}-\alpha  \left( \dot{x}_{n}+\frac{\alpha  \gamma }{
2 \eta }-\alpha  x_{n}\right)  \nonumber\\
\hphantom{\ddot{x}_{n}=}{}
 +(\dot{x}_{n}-\alpha  x_{n}) \sum_{\ell =1,\, \ell \neq n}^{N}\left[ \frac{
 ( \dot{x}_{\ell }-\alpha  x_{\ell } )   ( x_{n}+x_{\ell
} ) }{( x_{n}-x_{\ell })  x_{\ell }}\right]  ;
\end{gather}

(5) $b=\gamma =0$,   $\alpha =\rho =-2 a \eta$,   $f_{n}=-\alpha  x_{n}$, $\lambda =0$, $\mu =\alpha $:
\begin{gather}
 \ddot{x}_{n}=\frac{\dot{x}_{n}^{2}}{x_{n}}+\beta  \dot{x}_{n} x_{n}+\alpha
 \left( \dot{x}_{n}-\alpha  x_{n}-\beta  x_{n}^{2}\right)  \nonumber\\
 \hphantom{\ddot{x}_{n}=}{}
 + ( \dot{x}_{n}-\alpha  x_{n} )  \sum_{\ell =1,\, \ell \neq n}^{N}
\left[ \frac{ ( \dot{x}_{\ell }-\alpha  x_{\ell } )   (
x_{n}+x_{\ell } ) }{ ( x_{n}-x_{\ell } )  x_{\ell }}\right]  ;
\end{gather}

(6) $b=0$,   $\alpha =2 a \eta$,   $\beta =4 a \eta ^{2}/\gamma =2 \eta  \alpha
/\gamma$,   $\rho =-\alpha$, $ f_{n}=\alpha  z_{n}$, $  \lambda =0$, $ \mu =-\alpha $:
\begin{gather}
 \ddot{x}_{n}=\frac{\dot{x}_{n}^{2}}{x_{n}}+\frac{2 \alpha  \eta }{\gamma } %
\dot{x}_{n} x_{n}-\alpha  \left( \dot{x}_{n}+\alpha  x_{n}-\frac{2 \alpha
 \eta }{\gamma } x_{n}^{2}\right)    \nonumber \\
\hphantom{\ddot{x}_{n}=}{}
+ ( \dot{x}_{n}+\alpha  x_{n} )  \sum_{\ell =1,\, \ell \neq n}^{N}
\left[ \frac{ ( \dot{x}_{\ell }+\alpha  x_{\ell } )   (
x_{n}+x_{\ell } ) }{ ( x_{n}-x_{\ell } )  x_{\ell }}\right]  ;
\end{gather}

(7) $a=0$,  $ \alpha =-b \gamma$,  $ \beta =-2 b \eta$,   $f_{n}=\alpha
 x_{n}-\beta  x_{n}^{2}$, $\lambda =0$, $\mu =-\alpha $:
\begin{gather}
 \ddot{x}_{n}=\frac{\dot{x}_{n}^{2}}{x_{n}}+\beta  \dot{x}_{n} x_{n}+\rho
 \left( \dot{x}_{n}+\alpha  x_{n}+\beta  x_{n}^{2}\right)  \nonumber \\
\hphantom{\ddot{x}_{n}=}{}
 +\left( \dot{x}_{n}+\alpha  x_{n}-\beta  x_{n}^{2}\right)
 \sum_{\ell =1,\, \ell \neq n}^{N}\left[ \frac{\left( \dot{x}_{\ell
}+\alpha  x_{\ell }-\beta  x_{\ell }^{2}\right)   ( x_{n}+x_{\ell
} ) }{ ( x_{n}-x_{\ell } )  x_{\ell }}\right]  ;
\end{gather}

(8) $a=0$,   $\beta =-2 b \eta$,   $\rho =-b \gamma$, $ f_{n}=-b \gamma
 x_{n}+2 b \eta  x_{n}^{2}$, $ \lambda =(\alpha + b \gamma ) \gamma
/ ( 2 \eta  )$, $\mu =\alpha +2 b\gamma $:
\begin{gather}
 \ddot{x}_{n}=\frac{\dot{x}_{n}^{2}}{x_{n}}+\frac{\left( \alpha + b \gamma
\right)  \gamma }{2 \eta } \frac{\dot{x}_{n}}{x_{n}}-2 b \eta  \dot{x}
_{n} x_{n}
 -b \gamma  \left[ \dot{x}_{n}+\frac{ ( \alpha +b \gamma  )  \gamma
}{2 \eta }- ( \alpha +2 b \gamma  )  x_{n}+ 2 b \eta  x_{n}^{2}
\right]    \nonumber \\
\hphantom{\ddot{x}_{n}=}{}
 +\left( \dot{x}_{n}-b \gamma  x_{n}+2 b \eta  x_{n}^{2}\right)
 \sum_{\ell =1,\, \ell \neq n}^{N}\left[ \frac{\left( \dot{x}_{\ell
}-b \gamma  x_{\ell }+2 b \eta  x_{\ell }^{2}\right)   ( x_{n}+x_{\ell
} ) }{(x_{n}-x_{\ell }) x_{\ell }}\right]  ;
\end{gather}

(9) $\gamma =\rho =0$,   $\beta =2 b \eta$,   $f_{n}=2 a \eta  x_{n}+2 b \eta
 x_{n}^{2}$,   $\lambda =0$,   $\mu =\alpha $:
\begin{gather}
 \ddot{x}_{n}=\frac{\dot{x}_{n}^{2}}{x_{n}}-2 b \eta  \dot{x}_{n} x_{n}
+\left( \dot{x}_{n}+2 a \eta  x_{n}+2 b \eta  x_{n}^{2}\right)  \nonumber\\
\hphantom{\ddot{x}_{n}=}{}\times
 \sum_{\ell =1,\, \ell \neq n}^{N}\left[ \frac{\left( \dot{x}_{\ell
}+ 2 a \eta  x_{\ell }+ 2 b \eta  x_{\ell }^{2}\right)   ( x_{n}+x_{\ell
} ) }{ ( x_{n}-x_{\ell } )  x_{\ell }}\right]  ;
\end{gather}

(10) $\gamma =0$,   $\alpha =-2 a \eta$,   $\beta =-2 b \eta$,   $f_{n}=-\alpha
 x_{n}-\beta  x_{n}^{2}$,   $\lambda =0$,   $\mu =\alpha $:
\begin{gather}
 \ddot{x}_{n}=\frac{\dot{x}_{n}^{2}}{x_{n}}+\beta  \dot{x}_{n} x_{n}+\rho
 \left( \dot{x}_{n}-\alpha  x_{n}-\beta  x_{n}^{2}\right)    \nonumber \\
 \hphantom{\ddot{x}_{n}=}{}
 +\left( \dot{x}_{n}-\alpha  x_{n}-\beta  x_{n}^{2}\right)  \sum_{\ell
=1,\, \ell \neq n}^{N}\left[ \frac{\left( \dot{x}_{\ell }-\alpha  x_{\ell
}-\beta  x_{\ell }^{2}\right)   ( x_{n}+x_{\ell } ) }{ (
x_{n}-x_{\ell } )  x_{\ell }}\right]  ;
\end{gather}

(11) $a=-b \gamma$,  $\beta =-2 b \eta$,   $\rho =-\alpha -b \gamma$, $f_{n}=- ( 2 \eta +1 )  b \gamma  x_{n}+2 b \eta  x_{n}^{2}$, $\lambda = ( \alpha +b \gamma )  \gamma / ( 2 \eta  )$,
  $\mu =\alpha +2 b \gamma $:
\begin{gather}
 \ddot{x}_{n}=\frac{\dot{x}_{n}^{2}}{x_{n}}+\frac{ ( \alpha +b \gamma
 )  \gamma }{2 \eta } \frac{\dot{x}_{n}}{x_{n}}-2 b \eta  \dot{x}%
_{n} x_{n} - ( \alpha +b \gamma  )  \left[ \dot{x}_{n}+\frac{ ( \alpha
+b \gamma  )  \gamma }{2 \eta }- ( \alpha + 2 b \gamma  )
 x_{n}+2 b \eta  x_{n}^{2}\right]    \nonumber \\
 \hphantom{\ddot{x}_{n}=}{}
 +\left[ \dot{x}_{n}- ( 2 \eta +1 )  b \gamma  x_{n}+2 b \eta
 x_{n}^{2}\right] \nonumber\\
 \hphantom{\ddot{x}_{n}=}{}\times
  \sum_{\ell =1,\, \ell \neq n}^{N}\left\{ \frac{\left[ \dot{x}_{\ell
}- ( 2 \eta +1 )  b \gamma  x_{\ell }+2 b \eta  x_{\ell }^{2}\right]
 ( x_{n}+x_{\ell } ) }{ ( x_{n}-x_{\ell } )  x_{\ell }}
\right\}  ;
\end{gather}

(12) $\alpha =-2 a \eta$, $  \beta =-2 b \eta$, $  \rho =-\alpha
-b \gamma$, $ f_{n}= ( 2 a \eta -b \gamma  )  x_{n}+2 b \eta
 x_{n}^{2}$, $  \lambda = ( -2 a \eta + b \gamma  )  \gamma
/ ( 2 \eta  )$, $  \mu =-2 a \eta +2 b \gamma $:
\begin{gather}
 \ddot{x}_{n}=\frac{\dot{x}_{n}^{2}}{x_{n}}+\frac{ ( b \gamma -2 a \eta
 )  \gamma }{2 \eta } \frac{\dot{x}_{n}}{x_{n}}-2 b \eta  \dot{x}%
_{n} x_{n}  \nonumber \\
\hphantom{\ddot{x}_{n}=}{}
 - ( \alpha +b \gamma  )  \left[ \dot{x}_{n}+\frac{ ( b \gamma
- 2 a \eta  )  \gamma }{2 \eta }+ ( 2 a \eta -2 b \gamma  )
 x_{n}+2 b \eta  x_{n}^{2}\right]  \nonumber \\
\hphantom{\ddot{x}_{n}=}{}
+\left[ \dot{x}_{n}+ ( 2 a \eta -b \gamma  )  x_{n}+2 b \eta
 x_{n}^{2}\right]   \nonumber \\
\hphantom{\ddot{x}_{n}=}{}\times
  \sum_{\ell =1,\, \ell \neq n}^{N}\left\{ \frac{\left[ \dot{x}_{\ell
}+ ( 2 a \eta -b \gamma  )  x_{\ell }+2 b \eta  x_{\ell }^{2}\right]
  ( x_{n}+x_{\ell } ) }{ ( x_{n}-x_{\ell } )  x_{\ell }}%
\right\}  .
\end{gather}
\end{subequations}

$\eta \neq 0$,  case $(ii)$ ($5$ models, corresponding to Table~2.6):

(1) $a=b=0$,  $f_{n}=\lambda -\mu  x_{n}-\beta  x_{n}^{2}$, $ \lambda
= ( 2 \alpha  \eta -\beta  \gamma  )  \gamma / ( 4 \eta
^{2} )$, $\mu =\alpha -\beta  \gamma /\eta $:
\begin{subequations}
\begin{gather}
\ddot{x}_{n}=\frac{\dot{x}_{n}^{2}}{x_{n}}+\lambda  \frac{\dot{x}_{n}}{x_{n}%
}+\beta  \dot{x}_{n} x_{n}+\rho  \left( \dot{x}_{n}+\lambda -\mu
 x_{n}-\beta  x_{n}^{2}\right)   \nonumber \\
\hphantom{\ddot{x}_{n}=}{}
+\left( \dot{x}_{n}+\lambda -\mu  x_{n}-\beta  x_{n}^{2}\right)
 \sum_{\ell =1,\, \ell \neq n}^{N}\left[ \frac{\left( \dot{x}_{\ell
}+\lambda -\mu x_{\ell }-\beta  x_{\ell }^{2}\right)   ( x_{n}+x_{\ell
} ) }{( x_{n}-x_{\ell } )  x_{\ell }}\right]  ;
\end{gather}

(2) $b=\rho =0$,   $f_{n}=\lambda + ( 2 a \eta -\mu  )  x_{n}-\beta
 x_{n}^{2}$, $\lambda = ( 2 \alpha  \eta -\beta  \gamma  )
 \gamma / ( 4 \eta ^{2} )$, $ \mu =\alpha -\beta  \gamma /\eta $:
\begin{gather}
\ddot{x}_{n}=\frac{\dot{x}_{n}^{2}}{x_{n}}+\lambda  \frac{\dot{x}_{n}}{x_{n}}%
+\beta  \dot{x}_{n} x_{n}
+\left[ \dot{x}_{n}+\lambda + ( 2 a \eta -\mu  )  x_{n}-\beta
 x_{n}^{2}\right]    \nonumber \\
 \hphantom{\ddot{x}_{n}=}{}
  \sum_{\ell =1,\, \ell \neq n}^{N}\left\{ \frac{\left[\dot{x}_{\ell }+\lambda
+ ( 2 a \eta -\mu  )  x_{\ell }-\beta  x_{\ell }^{2}\right]  (
x_{n}+x_{\ell } ) }{ ( x_{n}-x_{\ell } )  x_{\ell }}\right\}
 ;
\end{gather}

(3) $a=0$,   $\alpha =b \gamma +\rho$,   $\beta =2 b \eta$,   $f_{n}=\lambda
-\rho  x_{n}$, $\lambda = ( \rho -b \gamma  )  \gamma / (
2 \eta  )$, $  \mu =\rho -b \gamma $:
\begin{gather}
 \ddot{x}_{n}=\frac{\dot{x}_{n}^{2}}{x_{n}}+\lambda \frac{\dot{x}_{n}}{x_{n}}%
+2 b \eta  \dot{x}_{n} x_{n}
 +\rho  \left[ \dot{x}_{n}+\lambda- ( \rho -b \gamma  )
 x_{n}-2 b \eta  x_{n}^{2}\right]    \nonumber \\
 \hphantom{\ddot{x}_{n}=}{}
+ ( \dot{x}_{n}+\lambda -\rho  x_{n} )
  \sum_{\ell =1,\, \ell \neq n}^{N}\left[ \frac{ ( \dot{x}_{\ell
}+\lambda -\rho x_{\ell } )   ( x_{n}+x_{\ell } ) }{(x_{n}-x_{\ell })  x_{\ell }}\right]  ;
\end{gather}

(4) $a=\gamma =0$,   $\alpha =\rho$,   $\beta =2 b \eta$,   $f_{n}=-\alpha
 x_{n}$,   $\lambda =0$,   $\mu =\alpha $:
\begin{gather}
 \ddot{x}_{n}=\frac{\dot{x}_{n}^{2}}{x_{n}}+\beta  \dot{x}_{n} x_{n}+\alpha
 \left( \dot{x}_{n}-\alpha  x_{n}-\beta  x_{n}^{2}\right)  \nonumber \\
\hphantom{\ddot{x}_{n}=}{}
+ ( \dot{x}_{n}-\alpha  x_{n} )  \sum_{\ell =1,\, \ell \neq n}^{N}%
\left[ \frac{ ( \dot{x}_{\ell }-\alpha  x_{\ell } )  (
x_{n}+x_{\ell } ) }{ ( x_{n}-x_{\ell } )  x_{\ell }}\right]  ;
\end{gather}

(5) $\alpha =2 a \eta$,   $\beta =2 b \eta$,   $\rho =-b \gamma
$,   $f_{n}=\lambda +b \gamma  x_{n}$, $  \lambda = ( 2 \alpha  \eta
-\beta  \gamma  )  \gamma / ( 2 \eta )$, $  \mu =2 a \eta
-2 b \gamma $:
\begin{gather}
 \ddot{x}_{n}=\frac{\dot{x}_{n}^{2}}{x_{n}}+\lambda \frac{\dot{x}_{n}}{x_{n}}%
+2 b \eta  \dot{x}_{n} x_{n}
 -b \gamma  \left[ \dot{x}_{n}+\lambda- ( 2 \alpha  \eta -\beta  \gamma
 )  x_{n}-2 b \eta  x_{n}^{2}\right]    \nonumber \\
 \hphantom{\ddot{x}_{n}=}{}
 + ( \dot{x}_{n}+\lambda+b \gamma  x_{n} )  \sum_{\ell =1,\ell
\neq n}^{N}\left[ \frac{ ( \dot{x}_{\ell }+\lambda +b \gamma  x_{\ell
} )   ( x_{n}+x_{\ell } ) }{ ( x_{n}-x_{\ell } )
 x_{\ell }}\right]  .
\end{gather}
\end{subequations}

\section{Second appendix}\label{appendixB}

In this appendix we list the second series of $24$ \textit{Newtonian}
equations of motion whose \textit{solvable} character has been demonstrated
in this paper; they correspond to those reported in Appendix~\ref{appendixA} via the
transformation among the $N$ zeros $z_{n}$ and the $N$ coef\/f\/icients $c_{m}$
of a monic polynomial, see~(\ref{Psi}) and~(\ref{chi}). In each case the
parameters they feature (such as $a$, $b$, $\alpha$, $\beta$, $\gamma$, $\eta$, $\rho $, as the case may be) are \textit{arbitrary constants}; the
(assigned) values of the other ones of these parameters (which also
characterize the time-evolution of the solutions of these equations, see~(\ref{ExplSolUt})), are also reported. Let us emphasize that if the parameter
$\rho $ is an \textit{imaginary} number and the parameter~$\alpha $ is
\textit{real} and \textit{negative}, the corresponding many-body problem is
\textit{asymptotically isochronous} with period $T$, see  Remark~\ref{remark2.3}; and that \textit{isochronous} many-body models are characterized by the $4$
\textit{Newtonian} equations of motion~(\ref{Isocm}), (\ref{Isocm1}) and~(\ref{Isocm2}) displayed at the end of Subsection~\ref{section2.3}. Let us also mention
again that the equations of motion reported below are \textit{not} all new;
in particular all those that are \textit{linear} are of course \textit{not}
new.

Let us recall that it is always assumed that $c_{n}=0$ for $n<0$ and for $n>N$, and $c_{0}=1 $.

$\eta =0$,  case $(i)$ ($5$ models, corresponding to  Table~2.3):

(1) $a=\beta =0$, $\alpha =-b \gamma$:
\begin{subequations}
\begin{gather}
\ddot{c}_{m}+[(1-2m) \alpha -\rho ] \dot{c}_{m}+m \alpha  [(m-1) \alpha
+\rho ] c_{m}=0 ;
\end{gather}

(2) $\beta =\rho =0$, $\alpha =-b \gamma$:
\begin{gather}
\ddot{c}_{m}-2(N-m-1) a \gamma  \dot{c}_{m-1}+(2 m-1) b \gamma  \dot{c}_{m}
+(N-m+2) (N-m+1) a^{2} \gamma ^{2} c_{m-2}
  \nonumber \\
\qquad{}
-2(N-m+1) (m-1) a b \gamma ^{2} c_{m-1}+m (m-1) b^{2} \gamma ^{2}=0;
\end{gather}

(3) $\alpha =\beta =0$, $\rho =-b \gamma$:
\begin{gather}
\ddot{c}_{m}+[(1-2 m ) \rho -\alpha ] \dot{c}_{m}+m \rho  [(m-1) \rho
+\alpha ] c_{m}=0 ;
\end{gather}

(4) $a=0, \alpha =-b \gamma$, $\rho =2 \alpha$:
\begin{gather}
\ddot{c}_{m}-(2 m+1) \alpha  \dot{c}_{m}-2 \beta  \dot{c}_{m+1}
+[m (m+1) \alpha ^{2}-2 \alpha  \beta  c_{1}+2 \beta  \dot{c}_{1}] c_{m} \nonumber\\
\qquad{}
+2 (m+1) \alpha  \beta  c_{m+1}=0 ;
\end{gather}

(5) $\alpha =b \gamma$, $\beta =b^{2} \gamma  / a$, $\rho =-2 b \gamma$:
\begin{gather}
\ddot{c}_{m}-2 (N-m+1) a \gamma  \dot{c}_{m-1}+(2 m-1) b \gamma  \dot{c}_{m}-
\frac{2 b^{2} \gamma }{a} \dot{c}_{m+1}    \nonumber \\
\qquad{} +(N-m+2) (N-m+1) a^{2} \gamma ^{2} c_{m-2}
-2(m-1) (N-m+1) a b \gamma ^{2} c_{m-1}    \nonumber \\
\qquad{}
+\left[m (m-3) b^{2} \gamma ^{2}+\frac{2 b^{3} \gamma ^{2}}{a} c_{1}+\frac{%
2 b^{2} \gamma }{a} \dot{c}_{1}\right] c_{m}
-(m+1) \frac{2 b^{3} \gamma ^{2}}{a} c_{m+1}=0 .
\end{gather}
\end{subequations}

$\eta =0$,  case $(ii)$ ($2$ models, corresponding to  Table~2.4):

(1) $\alpha =\beta =0$, $\rho =-b \gamma$:
\begin{subequations}
\begin{gather}
\ddot{c}_{m}-2 (N-m+1) a \gamma  \dot{c}_{m-1}+(2 m-1) b \gamma  \dot{c}_{m}
+(N-m+2) (N-m+1) a^{2} \gamma ^{2} c_{m-2}  \nonumber \\
\qquad {} -2 (m-1) (N-m+1) a b \gamma ^{2} c_{m-1}+m (m-1) b^{2} \gamma ^{2} c_{m}=0 ;
\end{gather}

(2) $a=\beta =0, \alpha =b \gamma +\rho :$
\begin{gather}
\ddot{c}_{m}-(2 m \rho +b \gamma ) \dot{c}_{m}+\left[m b \gamma  \rho
-\left(N^{2}-m^{2}\right) \rho ^{2}\right] c_{m}=0 .
\end{gather}
\end{subequations}

$\eta \neq 0$,  case $(i)$ ($12$ models, corresponding to  Table~2.5):

(1) $a=b=\rho =0$, $\lambda = [ (2 \alpha  \eta -\beta  \gamma ) \gamma
 ]  /  ( 2 \eta  ) ^{2}$, $\mu =\alpha -\beta  \gamma  / \eta$:
\begin{subequations}
\begin{gather}
\ddot{c}_{m}-\lambda  \dot{c}_{m-1}-\frac{\dot{c}_{N}}{c_{N}} \dot{c}%
_{m}-\beta  \dot{c}_{m+1}+\lambda  \frac{\dot{c}_{N}}{c_{N}} c_{m-1}+\beta  %
\dot{c}_{1} c_{m}=0 ;
\end{gather}

(2) $b=\beta =\gamma =\rho =0$, $\lambda =0$, $\mu =\alpha$:
\begin{gather}
\ddot{c}_{m}-\left[(2 N-4 m) a \eta -\frac{\dot{c}_{N}}{c_{N}}\right] \dot{c}
_{m}-2 m a \eta  \left(2 a \eta +\frac{\dot{c}_{N}}{c_{N}}\right) c_{m}=0 ;
\end{gather}

(3) $b=\beta =\gamma =0$, $\alpha =-2 a \eta$, $\lambda =0$, $\mu =\alpha$:
\begin{gather}
\ddot{c}_{m}+\left[(N-2 m) \alpha -\rho -\frac{\dot{c}_{N}}{c_{N}}\right] \dot{c}_{m}
-\left[m (N-m) \alpha ^{2}+(N-m) \alpha  \rho -m \alpha  \frac{\dot{c}_{N}}{c_{N}} \right] c_{m}=0 ;
\end{gather}

(4) $b=\beta =0$, $\alpha =-2 a \eta$, $\rho =2 a \eta =-\alpha$, $\lambda
=-a \gamma$, $\mu =\alpha$:
\begin{gather}
\ddot{c}_{m}+a \gamma  \dot{c}_{m-1}-\left[(N-2 m+1) \alpha +\frac{\dot{c}_{N}}{%
c_{N}}\right] \dot{c}_{m}
+a \gamma  \left[(N-m-1) \alpha -\frac{\dot{c}_{N}}{c_{N}}\right] c_{m-1}    \nonumber
\\
\qquad{} +\left[-(N-m) \alpha ^{2}+2 (N-m) \alpha ^{2}+m \alpha  \frac{\dot{c}_{N}}{c_{N}}%
\right] c_{m}=0 ;
\end{gather}

(5) $b=\gamma =0$, $\alpha =\rho =-2 a \eta$, $\lambda =0$, $\mu =\alpha$:
\begin{gather}
\ddot{c}_{m}+\left[(N-2 m-1) \alpha -\frac{\dot{c}_{N}}{c_{N}}\right] \dot{c}_{m}
-\left[(m+1) (N-m) \alpha ^{2}+m \alpha  \frac{\dot{c}_{N}}{c_{N}}+\alpha  \beta
 c_{1}-\beta  \dot{c}_{1}\right] c_{m}    \nonumber \\
\qquad{} +(m+1) \alpha  \beta  c_{m+1}=0 ;
\end{gather}

(6) $b=0, \alpha =2 a \eta$, $\beta =4 a \eta ^{2} / \gamma$, $\rho =-2 a \eta$, $\lambda =0$, $\mu =-2 a \eta$:
\begin{gather}
\ddot{c}_{m}-\left[2(N-2 m+1) a \eta +\frac{\dot{c}_{N}}{c_{N}}\right] \dot{c}_{m}-
\frac{4 a \eta ^{2}}{\gamma } \dot{c}_{m+1}    \\
\qquad{} -2 a \eta  \left[2 (m-1) (N-m) a \eta +m \frac{\dot{c}_{N}}{c_{N}}-\frac{4 a \eta
^{2}}{\gamma } c_{1}-\frac{2 \eta }{\gamma } \dot{c}_{1}\right] c_{m}
-(m+1) \frac{8 a^{2} \eta ^{3}}{\gamma } c_{m+1}=0 ; \nonumber
\end{gather}

(7) $a=0, \alpha =-b \gamma$, $\beta =-2 b \eta$, $\lambda =0$, $\mu =-\alpha$:
\begin{gather}
\ddot{c}_{m}-\left[(N-2 m) \alpha +\rho +\beta  c_{1}+\frac{\dot{c}_{N}}{c_{N}}\right]
\dot{c}_{m}-(2 m+1) \beta  \dot{c}_{m+1}    \nonumber \\
\qquad {} -\left[m (N-m) \alpha ^{2}-(N-m) \alpha  \rho +(N-m) \alpha  \beta  c_{1}
+(\beta  c_{1}+m \alpha ) \frac{\dot{c}_{N}}{c_{N}}+\beta  \rho  c_{1}-\beta
 \dot{c}_{1}\right] c_{m}    \nonumber \\
\qquad{}
-\left\{(m+1) \beta  \rho -[2 N^{2}-(3 m+5) N+4 m+4] \alpha  \beta
+m \beta ^{2} c_{1}+(m+1) \beta  \frac{\dot{c}_{N}}{c_{N}}\right\} c_{m+1}
\nonumber \\
\qquad{}
-m (m+2) \beta ^{2} c_{m+2}=0 ;
\end{gather}

(8) $a=0$, $\beta =-2 b \eta$, $\rho =-b \gamma$, $\lambda =(\alpha +b \gamma
) \gamma  /  ( 2 \eta  )$, $\mu =\alpha +2 b \gamma$:
\begin{gather}
\ddot{c}_{m}-\lambda  \dot{c}_{m-1}+\left[(N-2 m+1) b \gamma -2 b \eta  c_{1}-%
\frac{\dot{c}_{N}}{c_{N}}\right] \dot{c}_{m}
+4 (m-1) \dot{c}_{m+1}\nonumber \\
\qquad +\lambda  \left[-(N-m+1) b \gamma +\frac{\dot{c}_{N}}{c_{N}}%
\right] c_{m-1}
-\bigg[(m+4) (N-m) b^{2} \gamma ^{2}+2 (2 N-2 m+1) b^{2} \gamma  \eta  c_{1}
\nonumber \\
\qquad{} -(N-m) b \alpha  \gamma +2 b \eta  \dot{c}_{1}+(m b \gamma -2 b \eta  c_{1}) %
\frac{\dot{c}_{N}}{c_{N}}\bigg] c_{m}    \nonumber \\
\qquad +2 \left[(m+1) (N-2 m+1) b^{2} \gamma  \eta
-2 m b^{2} \eta ^{2} c_{1}-(m+1) b \eta  \frac{\dot{c}_{N}}{c_{N}}\right] c_{m+1}\nonumber\\
\qquad{}+4 m (m+2) c_{m+2}=0 ;
\end{gather}

(9) $\gamma =\rho =0$, $\beta =-2 b \eta$, $\lambda =0$, $\mu =\alpha$:
\begin{gather}
\ddot{c}_{m}-\left[2 (N-2 m)+2 b \eta  c_{1}+\frac{\dot{c}_{N}}{c_{N}}\right] \dot{c}%
_{m}+2 (2 m+1) b \eta  \dot{c}_{m+1}    \nonumber \\
\qquad{} +\left[-4 m (N-m) a^{2} \eta ^{2}+4 (N-m) a b \eta ^{2} c_{1}+\beta  \dot{c}_{1}
+2 (b \eta  c_{1}-m a \eta ) \frac{\dot{c}_{N}}{c_{N}}\right] c_{m}    \nonumber
\\
\qquad {} -2 \left[2 (m+1) (N-2 m) a b \eta ^{2}+2 m b^{2} \eta ^{2} c_{1}+(m+1) b \eta  %
\frac{\dot{c}_{N}}{c_{N}}\right] c_{m+1}    \nonumber \\
\qquad{} +4 m (m+2) c_{m+2}=0 ;
\end{gather}

(10) $\gamma =0$, $\alpha =-2 a \eta$, $\beta =-2 b \eta$, $\lambda =0$, $\mu
=\alpha$:
\begin{gather}
\ddot{c}_{m}+\left[(N-2 m) \alpha -\rho +\beta  c_{1}-\frac{\dot{c}_{N}}{c_{N}}\right] %
\dot{c}_{m}-(2 m+1) \beta  \dot{c}_{m+1}   \nonumber \\
\qquad {} -\left[m (N-m) \alpha ^{2}+(N-m) \alpha  \rho -(N-m) \alpha  \beta  c_{1}+\beta
 \rho  c_{1}
-\beta  \dot{c}_{1}+(\beta c_{1}-m \alpha ) \frac{\dot{c}_{N}}{c_{N}}\right] c_{m}
   \nonumber \\
\qquad{} -\left[(m+1) (N-2 m) \alpha  \beta -(m+1) \beta  \rho +m \beta
^{2} c_{1}-(m+1) \beta  \frac{\dot{c}_{N}}{c_{N}}\right] c_{m+1}    \nonumber \\
\qquad {} +m (m+2) c_{m+2}=0 ;
\end{gather}

(11) $a=-b \gamma$, $\beta =-2 b \eta$, $\rho =-\alpha -b \gamma$, $\lambda
=(\alpha +b \gamma ) \gamma  /  ( 2 \eta  )$, $\mu =\alpha
+2 b \gamma$:
\begin{gather}
\ddot{c}_{m}-\lambda  \dot{c}_{m-1}
+\left\{[2 (N-2 m) \eta +N-2 m+1] b \gamma +\alpha -2 b \eta  c_{1}-\frac{\dot{c}%
_{N}}{c_{N}}\right\} \dot{c}_{m}    \nonumber \\
\qquad {} +2 (2 m+1) \dot{c}_{m+1}+\lambda  \left[\frac{\dot{c}_{N}}{c_{N}}-(N-m+1) (\alpha
+b \gamma )\right] c_{m-1}    \nonumber \\
\qquad{} -\bigg\{m (N-m) b^{2} \gamma ^{2} (2 \eta +1)^{2}\!-(N-m) (\alpha +b \gamma
) (\alpha +2 b \gamma ) +2 (N-m) b^{2} \gamma  \eta  (2 \eta +1) c_{1}  \nonumber \\
\qquad{}+2 b \eta  (\alpha +b \gamma
) c_{1}+2 b \eta  \dot{c}_{1}
-[2 b \eta  c_{1}+m b \gamma  (2 \eta +1)] \frac{\dot{c}_{N}}{c_{N}}\bigg\} c_{m}
  \nonumber \\
\qquad{} +2 \bigg[(m+1) (N-2 m) b^{2} \gamma  \eta  (2 \eta +1)+(m+1) b \eta  (\alpha
+b \gamma )
-2 m b^{2} \eta ^{2} c_{1}\nonumber\\
\qquad{} -(m+1) b \eta  \frac{\dot{c}_{N}}{c_{N}}%
\bigg] c_{m+1}+m (m+2) c_{m+2}=0 ;
\end{gather}

(12) $\alpha =-2 a \eta$, $\beta =-2 b \eta$, $\rho =-\alpha -b \gamma
=2 a \eta -b \gamma$, $\lambda =(b \gamma -2 a \eta ) \gamma  / \left( 2 \eta
\right)$, $\mu =2 b \gamma -2 a \eta$:
\begin{gather}
\ddot{c}-\lambda  \dot{c}_{m-1}-\left[(N-2 m+1) (2 a \eta -b \gamma )+2 b \eta
 c_{1}+\frac{\dot{c}_{N}}{c_{N}}\right] \dot{c}_{m}    \nonumber \\
\qquad{} +2 (2 m+1) b \eta  \dot{c}_{m+1}+\lambda  \left[(N-m-1) (2 a \eta -b \gamma )+%
\frac{\dot{c}_{N}}{c_{N}}\right] c_{m-1}    \nonumber \\
\qquad {} +\bigg\{-m (N-m) (2 a \eta -b \gamma )^{2}+2 (N-m+1) (2 a \eta -b \gamma ) b \eta
 c_{1}    \nonumber \\
\qquad{} +(N-m) (2 a \eta -b \gamma ) (2 a \eta -2 b \gamma )-2 b \eta  \dot{c}_{1}
+[2 b \eta  c_{1}-m (2 a \eta -b \gamma )] \frac{\dot{c}_{N}}{c_{N}}\bigg\} c_{m}
   \nonumber \\
\qquad{} -2 \left[(m+1) (N-2 m+1) (2 a \eta -b \gamma ) b \eta
+2 m b^{2} \eta ^{2} c_{1}+(m+1) b \eta  \frac{\dot{c}_{N}}{c_{N}}\right] c_{m+1}\nonumber\\
\qquad{}
 +4 m (m+2) b^{2} \eta ^{2} c_{m+2}=0 .
\end{gather}
\end{subequations}

$\eta \neq 0$,  case $(ii)$ ($5$ models, corresponding to Table~2.6):

(1) $a=b=0$, $\lambda = [ (2 \alpha  \eta -\beta  \gamma ) \gamma  ]
 /  ( 2 \eta  ) ^{2}$, $\mu =\alpha -\beta  \gamma  / \eta$:
\begin{subequations}
\begin{gather}
\ddot{c}_{m}-(2 N-2 m+1) \lambda  \dot{c}_{m-1}
+\left[(N-2 m) \mu -\rho +\beta  c_{1}+\lambda  \frac{c_{N-1}}{c_{N}}-\frac{\dot{c%
}_{N}}{c_{N}}\right] \dot{c}_{m}    \nonumber \\
\qquad{} -(2 m+1) \beta  \dot{c}_{m+1}+(N-m+2) (N-m) \lambda ^{2} c_{m-2}  -\bigg[(N-2 m) (N-m+1) \lambda  \mu
\nonumber \\
\qquad{} -(N-m+1) \rho  \lambda
+(N-m+1) \beta  \lambda  c_{1}+\lambda  \frac{\dot{c}_{N}}{c_{N}}-\lambda
^{2} \frac{c_{N-1}}{c_{N}}\bigg] c_{m-1}    \nonumber \\
\qquad{} +\bigg[m (N-m) \big(2 \beta  \lambda -\mu ^{2}\big)+(\beta  c_{1}-m \mu ) \left(\lambda  \frac{%
c_{N-1}}{c_{N}}-\frac{\dot{c}_{N}}{c_{N}}\right)    \nonumber \\
\qquad -(N-m) \rho  \mu +(N-m) \beta  \mu  c_{1}-\beta  \rho  c_{1}+\beta  \dot{c}%
_{1}\bigg] c_{m}   \nonumber \\
\qquad {} -\bigg[(m+1) (N-2 m) \beta  \mu +(m+1) \beta  \rho +m \beta
^{2} c_{1}-(m+1) \beta  \frac{\dot{c}_{N}}{c_{N}}    \nonumber \\
\qquad{} -(N-m-2) \beta  \lambda  \frac{c_{N-1}}{c_{N}}\bigg] c_{m+1}+m (m+2) \beta
^{2} c_{m+2}=0 ;
\end{gather}

(2) $b=\rho =0$, $\lambda = [ (2 \alpha  \eta -\beta  \gamma ) \gamma
 ]  /  ( 2 \eta  ) ^{2}$, $\mu =\alpha -\beta  \gamma  / \eta$:
\begin{gather}
\ddot{c}_{m}-(2 N-2 m+1) \lambda  \dot{c}_{m-1}
+\left[(N-2 m) (\mu -2 a \eta )+\beta  c_{1}+\lambda  \frac{c_{N-1}}{c_{N}}-\frac{%
\dot{c}_{N}}{c_{N}}\right] \dot{c}_{m}    \nonumber \\
\qquad{} -(2 m+1) \beta  \dot{c}_{m+1}+(N-m) (N-m+2) \lambda ^{2} c_{m-2} \nonumber\\
\qquad{}
+\bigg[(N-2 m) (N-m+1) \lambda  (2 a \eta -\mu )-(N-m+1) \beta  \lambda  c_{1}
\nonumber \\
\qquad{}+(N-m) \lambda  \frac{\dot{c}_{N}}{c_{N}}-(N-m) \lambda ^{2} \frac{c_{N-1}}{%
c_{N}}\bigg]c_{m-1}
+\bigg\{m (N-m) \big[2 \beta  \lambda -(2 a \eta -\mu )^{2}\big]\nonumber\\
\qquad {}-(N-m) (2 a \eta -\mu
) \beta  c_{1}+\beta  \dot{c}_{1}
+[\beta  c_{1}+m (2 a \eta -\mu )] \left(\lambda  \frac{c_{N-1}}{c_{N}}-\frac{%
\dot{c}_{N}}{c_{N}}\right)\bigg\} c_{m}   \nonumber \\
\qquad{} -\bigg[(m+1) (N-2 m) \beta  \mu +(m+1) \beta  \rho +m \beta ^{2} c_{1}
\nonumber \\
\qquad{}-(m+1) \beta  \frac{\dot{c}_{N}}{c_{N}}-(N-m-2) \beta  \lambda  \frac{c_{N-1}%
}{c_{N}}\bigg] c_{m+1}
+m (m+2) \beta ^{2} c_{m+2}=0 ;
\end{gather}

(3) $a=0$, $\alpha =b \gamma +\rho$, $\beta =2 b \eta$, $\lambda =\gamma  \rho
 /  ( 2 \eta  )$, $\mu =\rho -b \gamma$:
\begin{gather}
\ddot{c}_{m}-(2 N-2 m+1) \lambda  \dot{c}_{m-1}-\left[(N-2 m+1) \rho -\lambda  %
\frac{c_{N-1}}{c_{N}}+\frac{\dot{c}_{N}}{c_{N}}\right] \dot{c}_{m}    \nonumber \\
\qquad{} -2 b \eta  \dot{c}_{m+1}+(N-m) (N-m+2) \lambda ^{2} c_{m-2}    \nonumber \\
\qquad{} +\left[(N-m+1) (N-2 m+1) \lambda  \rho +(N-m) \lambda  \left(\frac{\dot{c}_{N}}{c_{N}}%
-\lambda  \frac{c_{N-1}}{c_{N}}\right)\right] c_{m-1}    \nonumber \\
\qquad {}-\bigg[(m+1) (N-m) \rho ^{2}-(N-m) b \gamma  \rho +m \rho  \frac{\dot{c}_{N}}{%
c_{N}}-m \lambda  \rho  \frac{c_{N-1}}{c_{N}}   \nonumber \\
\qquad{} -2 b \eta  \rho  c_{1}-2 b \eta  \dot{%
c}_{1}\bigg] c_{m}+2 (m+1) b \eta  \rho  c_{m+1}=0 ;
\end{gather}

(4) $a=\gamma =0$, $\alpha =\rho$, $\beta =2 b \eta$, $\lambda =0$, $\mu =\alpha$:
\begin{gather}
\ddot{c}_{m}+\left[(N-2 m-1) \alpha -\frac{\dot{c}_{N}}{c_{N}}\right] \dot{c}_{m}-\beta
 \dot{c}_{m+1}    \nonumber \\
\qquad {} -\left[(m+1) (N-m) \alpha ^{2}-m \alpha  \frac{\dot{c}_{N}}{c_{N}}+\alpha  \beta
 c_{1}-\beta  \dot{c}_{1}\right] c_{m}     +(m+1) \alpha  \beta  c_{m+1}=0 ;
\end{gather}

(5) $\alpha =2 a \eta$, $\beta =2 b \eta$, $\rho =-b \gamma$, $\lambda = [
(2 a \eta -b \gamma ) \gamma  ]  /  ( 2 \eta  )$, $\mu
=2 a \eta -2 b \gamma$:
\begin{gather}
\ddot{c}_{m}-(2 N-2 m+1) \lambda  \dot{c}_{m-1}-\left[(N-2 m-1) b \gamma -\lambda
 \frac{c_{N-1}}{c_{N}}+\frac{\dot{c}_{N}}{c_{N}}\right] \dot{c}_{m}    \nonumber
\\
\qquad{} -\left[(N+2 m+1) (N-m+1) b \gamma  \lambda +(N-m) \lambda  \left(\lambda  \frac{c_{N-1}%
}{c_{N}}-\frac{\dot{c}_{N}}{c_{N}}\right)\right] c_{m-1}    \nonumber \\
\qquad {} +\bigg[-(m+2) (N-m) b^{2} \gamma ^{2}+m b \gamma  \left(\lambda  \frac{c_{N-1}}{c_{N}}-%
\frac{\dot{c}_{N}}{c_{N}}\right)    \nonumber \\
\qquad{} +2 (N-m) a b \gamma  \eta +2 b^{2} \gamma  \eta  c_{1}+2 b \eta  \dot{c}%
_{1}\bigg] c_{m}
-2 (m+1) b^{2} \gamma  \eta  c_{m+1}=0 .
\end{gather}
\end{subequations}

\subsection*{Acknowledgements}

We would like to thank an unknown referee whose intervention allowed us to
eliminate a mistake contained in the original version of our paper.

\pdfbookmark[1]{References}{ref}
\LastPageEnding


\begin{thebibliography}{99}
\footnotesize\itemsep=0pt

\bibitem{OP1}
Olshanetsky M.A., Perelomov A.M., Explicit solution of the {C}alogero model in
  the classical case and geodesic f\/lows on symmetric spaces of zero curvature,
  \href{http://dx.doi.org/10.1007/BF02750226}{\textit{Lett. Nuovo Cimento}} \textbf{16} (1976), 333--339.

\bibitem{OP2}
Olshanetsky M.A., Perelomov A.M., Completely integrable {H}amiltonian systems
  connected with semisimple {L}ie algebras, \href{http://dx.doi.org/10.1007/BF01418964}{\textit{Invent. Math.}} \textbf{37}
  (1976), 93--108.


\bibitem{OP3}
Olshanetsky M.A., Perelomov A.M., Explicit solutions of some completely
  integrable systems, \href{http://dx.doi.org/10.1007/BF02720431}{\textit{Lett. Nuovo Cimento}} \textbf{17} (1976), 97--101.

\bibitem{OP4}
Olshanetsky M.A., Perelomov A.M., Classical integrable f\/inite-dimensional
  systems related to {L}ie algebras, \href{http://dx.doi.org/10.1016/0370-1573(81)90023-5}{\textit{Phys. Rep.}} \textbf{71} (1981),
  313--400.

\bibitem{C2001}
Calogero F., Classical many-body problems amenable to exact treatments,
  \textit{Lecture Notes in Physics. New Series~m: Monographs}, Vol.~66,
  Springer-Verlag, Berlin, 2001.

\bibitem{C2008}
Calogero F., Isochronous systems, \href{http://dx.doi.org/10.1093/acprof:oso/9780199535286.001.0001}{Oxford University Press}, Oxford, 2008.\footnote{A~paperback edition of this
monograph has been published by Oxford University Press in September~2012:
it coincides with the hardback version, except for the corrections of
several misprints, and the addition of a two-page ``Preface to the paperback
edition'' reporting a number of relevant new references.}


\bibitem{CC1}
Bruschi M., Calogero F., Droghei R., Integrability, analyticity, isochrony,
  equilibria, small oscillations, and {D}iophantine relations: results from the
  stationary {B}urgers hierarchy, \href{http://dx.doi.org/10.1088/1751-8113/42/47/475202}{\textit{J.~Phys.~A: Math. Theor.}} \textbf{42}
  (2009), 475202, 9~pages.

\bibitem{CC2}
Bruschi M., Calogero F., Droghei R., Integrability, analyticity, isochrony,
  equilibria, small oscillations, and {D}iophantine relations: results from the
  stationary {K}orteweg--de {V}ries hierarchy, \href{http://dx.doi.org/10.1063/1.3267067}{\textit{J.~Math. Phys.}}
  \textbf{50} (2009), 122701, 19~pages.

\bibitem{CC3}
Calogero F., A new class of solvable dynamical systems, \href{http://dx.doi.org/10.1063/1.2920569}{\textit{J.~Math. Phys.}}
  \textbf{49} (2008), 052701, 9~pages.

\bibitem{CC4}
Calogero F., Two new solvable dynamical systems of goldf\/ish type,
  \href{http://dx.doi.org/10.1142/S1402925110000970}{\textit{J.~Nonlinear Math. Phys.}} \textbf{17} (2010), 397--414.

\bibitem{CC5}
Calogero F., Isochronous dynamical systems, \href{http://dx.doi.org/10.1098/rsta.2010.0250}{\textit{Philos. Trans. R. Soc.
  Lond. Ser. A Math. Phys. Eng. Sci.}} \textbf{369} (2011), 1118--1136.

\bibitem{CC6}
Calogero F., A new goldf\/ish model, \href{http://dx.doi.org/10.1007/s11232-011-0056-4}{\textit{Theoret. and Math. Phys.}}
  \textbf{167} (2011), 714--724.

\bibitem{CC7}
Calogero F., Another new goldf\/ish model, \href{http://dx.doi.org/10.1007/s11232-012-0060-3}{\textit{Theoret. and Math. Phys.}}
  \textbf{171} (2012), 629--640.

\bibitem{CC8}
Calogero F., An integrable many-body problem, \href{http://dx.doi.org/10.1063/1.3638052}{\textit{J.~Math. Phys.}}
  \textbf{52} (2011), 102702, 5~pages.


\bibitem{CC9}
Calogero F., New solvable many-body model of goldf\/ish type,
  \href{http://dx.doi.org/10.1142/S1402925112500064}{\textit{J.~Nonlinear Math. Phys.}} \textbf{19} (2012), 1250006, 19~pages.


\bibitem{CC10}
Calogero F., Two quite similar matrix {ODE}s and the many-body problems related
  to them, \href{http://dx.doi.org/10.1142/S021988781260002X}{\textit{Int.~J. Geom. Methods Mod. Phys.}} \textbf{9} (2012),
  1260002, 6~pages.

\bibitem{CC11}
Calogero F., Another new solvable many-body model of goldf\/ish type,
  \href{http://dx.doi.org/10.3842/SIGMA.2012.046}{\textit{SIGMA}} \textbf{8} (2012), 046, 17 pages, \href{http://arxiv.org/abs/1207.4850}{arXiv:1207.4850}.




\bibitem{CC12}
Calogero F., Iona S., Isochronous dynamical system and {D}iophantine
  relations.~{I}, \href{http://dx.doi.org/10.1142/S1402925109000091}{\textit{J.~Nonlinear Math. Phys.}} \textbf{16} (2009),
  105--116.

\bibitem{CC13}
Calogero F., Leyvraz F., Examples of isochronous {H}amiltonians: classical and
  quantal treatments, \href{http://dx.doi.org/10.1088/1751-8113/41/17/175202}{\textit{J.~Phys.~A: Math. Theor.}} \textbf{41} (2008),
  175202, 11~pages.

\bibitem{CC14}
Calogero F., Leyvraz F., A new class of isochronous dynamical systems,
  \href{http://dx.doi.org/10.1088/1751-8113/41/29/295101}{\textit{J.~Phys.~A: Math. Theor.}} \textbf{41} (2008), 295101, 14~pages.

\bibitem{CC15}
Calogero F., Leyvraz F., Isochronous oscillators, \href{http://dx.doi.org/10.1142/S1402925110000611}{\textit{J.~Nonlinear Math.
  Phys.}} \textbf{17} (2010), 103--110.

\bibitem{CC16}
Calogero F., Leyvraz F., Solvable systems of isochronous, multi-periodic or
  asymptotically isochronous nonlinear oscillators, \href{http://dx.doi.org/10.1142/S1402925110000623}{\textit{J.~Nonlinear Math.
  Phys.}} \textbf{17} (2010), 111--120.

\bibitem{CC17}
Droghei R., Calogero F., Ragnisco O., An isochronous variant of the
  {R}uijsenaars--{T}oda model: equilibrium conf\/igurations, behavior in their
  neighborhood, {D}iophantine relations, \href{http://dx.doi.org/10.1088/1751-8113/42/44/445207}{\textit{J.~Phys.~A: Math. Theor.}}
  \textbf{42} (2009), 445207, 9~pages.

\bibitem{HTF1}
Erd\'elyi A., Magnus W., Oberhettinger F., Tricomi F.G. (Editors), Higher
  transcendental functions, Vol.~I, \textit{Bateman Manuscript Project}, McGraw-Hill
  Book Co., New York, 1953.

\end{thebibliography}
\end{document}